\documentclass[aps,twocolumn,preprintnumbers,amsmath,amssymb,nofootinbib,superscriptaddress,notitlepage]{revtex4-1}

\usepackage{epsfig}
\usepackage[utf8]{inputenc}

\begin{document}

\title{Renormalization group evolution and power counting in nuclear matter}

\author{Manuel Pavon Valderrama}\email{mpavon@buaa.edu.cn}
\affiliation{School of Physics, Beihang University, Beijing 100191, China} 

\date{\today}


\begin{abstract} 
  In nuclear matter, for interparticle separations larger than the healing
  distance (a characteristic long-distance scale of finite-density fermionic
  systems), the in-medium two-body wave function is essentially
  a free wave function.
  In terms of the renormalization group (RG), this implies that the running of
  the effective field theory (EFT) couplings freezes for r-space cutoffs
  above this distance (or p-space cutoffs below the corresponding
  healing momentum scale).
  As a consequence the leading order EFT description of nuclear matter
  (understood here as the infrared limit of the RG) corresponds to
  the mean-field approximation and a set of tree-level
  (i.e. perturbative) leading contact-range couplings.
  Though the contacts do in principle inherit the power counting
  they had in the vacuum, their iteration is suppressed
  in the infrared, explaining why they become
  perturbative in nuclear matter.
  In addition to the contact-range potential, RG evolution requires
  the inclusion of density-dependent terms in the equation of state
  that can be represented by a pseudo-potential, which
  (unlike the genuine contacts) should not be iterated.
  The ${\rm LO}$ EFT description ends up being a subset of Skyrme forces
  previously identified by the Orsay group.
\end{abstract}

\maketitle

\section{Introduction}

Effective field theories (EFTs) are model independent and systematically
improvable descriptions of low energy phenomena~\cite{Georgi:1993mps}.
They are ideally suited for situations in which a more fundamental,
high energy description is unavailable or simply impractical.
The fundamental theory of strong interactions is quantum chromodynamics (QCD),
which owing to asymptotic freedom is not analytically solvable at
the characteristic low momenta of nucleons within nuclei.
EFTs are the natural answer to this limitation~\cite{Epelbaum:2008ga,Machleidt:2011zz,Hammer:2019poc,vanKolck:2020llt,Machleidt:2024bwl},
which otherwise can only be overcome by brute force,
i.e., lattice QCD~\cite{Beane:2010em,Aoki:2012tk}.
In addition, they are able to bridge the gap between QCD and nuclear physics
thanks to the fact that EFTs can be understood as the infrared
renormalization group (RG) evolution of their underlying
theory (QCD in this case).

This connection is well-understood regarding the derivation of
nuclear forces, where the EFT potential takes the form of
a power series in terms of the small ratio $Q/M$:
\begin{eqnarray}
  V_{\rm EFT} = \sum_{\nu} V^{(\nu)}\,{\left( \frac{Q}{M} \right)}^{\nu} \, ,
\end{eqnarray}
with $Q$ and $M$ referring, respectively, to the low- and high-energy
scales of the theory. In nuclear physics $Q$ might represent the
pion mass $m_{\pi}$ or the momentum exchanged by the nucleons
$|\vec{q}\,|$, while $M$ refers to the rho or
nucleon masses, $m_{\rho}$ or $m_N$.
The set of rules by which this expansion is structured is
the power counting of the EFT, which is in turn
intimately intertwined with its infrared RG
flow~\cite{Birse:1998dk,Barford:2002je}.

The extension of the EFT description to finite nuclei and infinite
nuclear matter is conceptually straightforward, but often
unachievable in practice.
While in principle it merely entails the direct calculation of nuclear
properties from few-nucleon forces, the limitation remains that this
type of ab-initio approach is only computationally viable
for light- and medium-mass nuclei.

Heavy-mass nuclei and nuclear matter are often described instead
in terms of effective interactions (effective, that is, in the
phenomenological sense) such as Skyrme~\cite{Skyrme:1959zz}
or Gogny~\cite{Decharge:1979fa}.
However, they often lack a clear connection to an underlying Hamiltonian
or Lagrangian formulation~\cite{Duguet:2013dga,Grasso:2018pen,Colo:2020vik},
which prevents us from deriving them within the EFT frameworks
used to describe few-nucleon systems.
In principle, these systems are still suspected to be amenable to an EFT
description of their own --- energy density functionals (EDFs) and
the aforementioned phenomenological effective interactions do have
a strong EFT resemblance or flavor~\cite{Furnstahl:2007xm,Drut:2009ce,Furnstahl:2019lue},
so to speak ---, though current efforts to formulate a rigorous EFT
for EDFs remain incomplete despite
ongoing efforts~\cite{Kaiser:2015vpa,Lacroix:2016dfs,Yang:2013bhd,Yang:2016mvq,Yang:2017shi,Burrello:2020myg,Yang:2021akb,Gebremariam:2010ni,NavarroPerez:2018tme,Bulgac:2017bho,Zurek:2023mdh}.

Here I will consider the RG evolution of EFT contact-range
interactions in nuclear matter.
The key idea is the identification of the healing distance~\cite{Brueckner:1955zzb,Brueckner:1955zza,Bethe:1957aa,Gomes:1957zz}
as the relevant scale for defining the infrared limit of
strongly interacting systems of fermions at finite density.
This is the distance above which the two-nucleon wave function in the medium
essentially becomes a plane wave.
It is a consequence of the Pauli exclusion principle, which strongly
suppresses scattering between nucleon pairs in the Fermi sea,
resulting in a free wave function (modulo small
oscillatory corrections) for momenta below
the Fermi momentum.
The RG evolution of the contact-range couplings depends on the form of
the wave function (as explicitly shown in~\cite{Valderrama:2014vra}),
where for a free wave function the couplings are approximately
scale independent.
That is, the RG flow of the couplings basically freezes at scales softer
than the healing distance, while at harder scales it will be given by
two-body correlations (i.e. the in-medium two-body wave function)
and reproduce the vacuum RG flow.
Moreover, as is well-known, the in-medium two-body phase shifts
become zero~\cite{Bethe:1957aa}, i.e., in-medium two-body scattering is trivial.

In terms of the RG and its infrared evolution, the previous observation
suggests the formulation of a {\it perturbative} low energy EFT
for nuclear matter.
It is perturbative in the sense that interactions are not to be iterated at
all orders (as implied by the triviality of in-medium two-body scattering),
despite the fact that the size of said interactions is still
infrared-enhanced.
This non-perturbative to perturbative transition is not
trivial to prove though.
The lowest order of this EFT corresponds to the mean-field (MF) approximation,
in which couplings are approximately scale independent, while the vacuum
running of these couplings at harder scales is recovered
order-by-order with beyond-mean-field (BMF) corrections.
This second observation defines the conditions by which the low energy
description of nuclear matter can be connected with
that of few-nucleon forces (in particular,
their contact-range components).

The manuscript is structured as follows: in Section~\ref{sec:healing}
I illustrate with the concrete example of a contact-range theory
the appearance of the healing distance, how it leads to
the freeze of the RG evolution of the couplings and
how loop corrections become perturbative.
In Section~\ref{sec:nuclear-matter} I analyze the power counting of
contact-range two-body forces in nuclear matter, and briefly
discuss the role of pion exchanges and few-body forces.
Section~\ref{sec:density-dependence} deals with the appearance of
density-dependent contributions to the energy per nucleon that
originate specifically from the infrared RG evolution
in nuclear matter.
Finally, in Section~\ref{sec:discussion} I summarize and discuss
the main findings of the present work.
Appendices \ref{app:g-matrix}, \ref{app:loop} and \ref{app:counting}
include technical details or in-depth discussions of the G-matrix,
the calculation of loop corrections in nuclear matter, and
power counting, respectively.

\section{The healing distance and the renormalization group flow}
\label{sec:healing}

One interesting characteristic of infinite fermion systems is that
the two-body wave function quickly approaches the free wave
function at large interparticle distances.
As a consequence of the exclusion principle, the fermions below the Fermi
momentum (i.e., at low energies) are prevented from scattering
with each other and thus end up behaving as free particles~\cite{Bethe:1957aa}.

In nuclear physics this is referred to as the healing of
the wave function~\cite{Gomes:1957zz}.
From the EFT perspective, this has very interesting consequences because
the running of the contact-range couplings can be determined
from the behavior of the wave function at different
distances or momenta~\cite{Valderrama:2014vra}.
For a perturbative contact-range coupling, its running is determined by
\begin{eqnarray}
  \frac{d}{d \Lambda}\, \Big[ C(\Lambda)\, \langle \Psi_{\rm EFT} |
    \mathcal{O}_C | \Psi_{\rm EFT} \rangle \Big] = 0 \, ,
\end{eqnarray}
where $\Psi_{\rm EFT}$ is the EFT wave function, $\Lambda$ the cutoff,
$\mathcal{O}_C$ a contact-range operator (i.e. a regularized
Dirac-delta) and $C(\Lambda)$ the coupling.
If $\Psi_{\rm EFT}$ is a free wave function, the RG equation simplifies to
\begin{eqnarray}
  \frac{d}{d \Lambda} C(\Lambda) = 0 \, ,
\end{eqnarray}
which implies that the coupling does not run.
This happens independently of whether one originally relies on perturbative or
non-perturbative arguments, though if the two-body wave function is
well approximated by a plane wave at large distances is because
the contact-range interaction is indeed perturbative
at low energies.

For nuclear matter the previous argument implies that the evolution of the
coupling constants approximately freezes for cutoffs softer
than the healing distance.
Thus the infrared limit of the RG equations results in couplings that
eventually become independent of the cutoff, that is
\begin{eqnarray}
  C(\Lambda < \Lambda^*) \approx C(\Lambda^*) \, ,
\end{eqnarray}
where $\Lambda^*$ is a momentum scale that is related to the healing
distance or the Fermi momentum.
If non-running couplings are indeed perturbative (independently of the
reason why they do not run), this leads to a very convenient and
simple EFT description of nuclear matter in which the contact-range
interactions are perturbative (regardless of whether this is the case
in the vacuum) and their size given by the new momentum
scale $\Lambda^*$.
Incidentally, this is how Skyrme interactions work (for which
calculations are usually done at tree level).

It is interesting to notice that if one is using the MF approximation,
the two-body wave function is indeed a free wave function.
That is, in this approximation the coupling constants
do not run and are cutoff independent instead.
This also implies that if one wants to recover in the medium the ultraviolet
running of the coupling constants in the vacuum (i.e. the running for
$\Lambda > \Lambda^*$) it will be necessary to include BMF
corrections so as to reproduce the two-body short-range
correlations on which the ultraviolet
running depends.

The practical consequence of this observation is that
if one wants to {\it directly}~\footnote{The stress here is on directly.
  Indirect connections~\cite{Bogner:2003wn,Bogner:2006pc,Bogner:2014baa,Binder:2015trg} (e.g. descriptions in a reduced or tailored model space
  for the extended system)
  are in general not limited by the same constraints as a direct
  matching with nuclear EFT.} connect the EFT description of
few-nucleon systems in the vacuum with that of extended systems, such as
nuclear matter or heavy nuclei, then it is necessary to concentrate
on soft cutoffs.
Descriptions with hard cutoffs instead require BMF corrections in order to
recover (in the form of an expansion if MF is the starting point)
the renormalizability properties of the EFT in the vacuum.

The rest of this section explains in more detail the healing of the wave
function and how it explicitly affects the running of
the coupling constant.
For this, (i) the vacuum and in-medium two-body wave functions are compared,
(ii) a cutoff is introduced and from it the running of the lowest order
coupling constant is calculated and (iii) the power counting of this
coupling, in particular whether it is perturbative or not,
is discussed.
Basically, I will be giving a more detailed derivation of the physical
intuitions discussed in the previous paragraphs.
The formalism relies on the r-space representation of the Lippmann-Schwinger
and Bethe-Goldstone equations.
Not every EFT practitioner is familiar with this choice, though this is not
a problem: the arguments about power counting will be derived again
in Section \ref{sec:nuclear-matter} in a more usual language.
Further details about the calculations appearing in the present Section
are provided in Appendix~\ref{app:g-matrix}.

\subsection{The two-body wave function}

In the vacuum, the zero-energy S-wave wave function of
a two-body system with a zero-range interaction
is given by the well-known expression
\begin{eqnarray}
  \Psi_0(\vec{r}\,) =
  \frac{1}{\sqrt{4\pi}}\frac{u_0(r)}{r} \propto
  \left( 1 - \frac{a_0}{r} \right) \, ,
\end{eqnarray}
where $r$ is the radius and $a_0$ the scattering length.
The intention here is to compare this wave function with the corresponding
one for a two-fermion system below the Fermi sea.
Even though in this latter case analytic expressions
are known (actually for a hard core~\cite{Dahll:1969hmo,Viollier:1976ab} or
a delta-shell~\cite{RuizSimo:2017tcb}, though they are easily
extended to the zero-range case), it would be convenient
nonetheless to explain how the in-medium
wave function is obtained.
This will prove useful for calculating the finite cutoff version of
this wave function and the running of the coupling constant at finite densities.

For deriving the in-medium two-body wave function it will be convenient to
consider first the following r-space version of the Lippmann-Schwinger
equation for the vacuum case
\begin{eqnarray}
  \Psi_{\vec{p}\,}(\vec{r}\,) = e^{i \vec{p} \cdot \vec{r}} +
  \int d^3\vec{r}\,'\,G_0(\vec{r} - \vec{r}\,')\,V(\vec{r}\,')\,
  \Psi_{\vec{p}\,}(\vec{r}\,') \, , \nonumber \\
\end{eqnarray}
in which $\vec{p}\,$ is the center-of-mass momentum of the two-body system and
$V$ the two-body potential (which is zero-range in the case at hand).
The propagator $G_0$ reads
\begin{eqnarray}
  G_0(\vec{r} - \vec{r}\,') &=& m_N\,\int \frac{d^3\vec{k}\,}{(2\pi)^3}\,
  \frac{e^{i \vec{k}\, \cdot (\vec{r} - \vec{r}\,')}}{p^2 + i \epsilon - k^2}
  \nonumber \\
  &=& -\frac{m_N}{4 \pi}\,\frac{e^{i p | \vec{r} - \vec{r}\,' |}}
            {|\vec{r} - \vec{r}\,'|} \, ,
\end{eqnarray}
and represents the propagation of the two particles within a loop, where
$m_N$ is the nucleon mass as I am assuming the particles to be nucleons.
Within this formalism there is a well-known extension for two fermions
interacting within the Fermi sea --- the Bethe-Goldstone
equation (check~\cite{Day:1967zza} for a review)---
which in r-space reads
\begin{eqnarray}
  \Psi_{\vec{p}\, , \vec{P}\,}(\vec{r}\,) = e^{i \vec{p} \cdot \vec{r}} +
  \int d^3\vec{r}\,'\,K_0(\vec{r} - \vec{r}\,')\,V(\vec{r}\,')\,
  \Psi_{\vec{p}\, , \vec{P}\,}(\vec{r}\,') \, , \nonumber \\
\end{eqnarray}
where the two-body wave function now depends not only on the relative momentum
$\vec{p}\, = (\vec{p}_1\, - \vec{p}_2\,)/2$ but also on
the total momentum $\vec{P} = \vec{p}_1 + \vec{p}_2\,$ (with $\vec{p}_1$ and
$\vec{p}_2$ the momenta of particles $1$ and $2$).
In this expression $K_0$ is the in-medium two-body propagator,
which reads 
\begin{eqnarray}
  && K_0(\vec{r} - \vec{r}\,') = \nonumber \\
  && \, m_N\,\int \frac{d^3\vec{k}\,}{(2\pi)^3}\,
  \frac{e^{i \vec{k}\, \cdot (\vec{r} - \vec{r}\,')}}{p^2 - k^2}\,
  \theta( k_1 - k_F)\,\theta( k_2 - k_F) \, ,
\end{eqnarray}
where $k_F$ is the Fermi momentum and with $k_1$, $k_2$ the magnitude of
the momenta of the two particles within the loop:
\begin{eqnarray}
  k_1 = | \frac{\vec{P}}{2}+\vec{k}\,| \quad \mbox{and} \quad
  k_2 = | \frac{\vec{P}}{2}-\vec{k}\,| \, .
\end{eqnarray}
That is, $K_0$ simply represents $G_0$ when rescattering is limited to
two-body states that are not within the Fermi sea (where
this explains why the in-medium two-body wave function depends
both on $\vec{p}\,$ and $\vec{P}\,$).

If one particularizes the previous expressions for a zero-range interaction
and for $\vec{p} = 0$ and $\vec{P} = 0$, one obtains
\begin{eqnarray}
  \Psi_0(\vec{r}\,) &=& 1 + \lambda\,G_0(r) \,  , \label{eq:2B-wf-vacuum} \\
  \Psi_{0,0}(\vec{r}\,) &=& 1 + \mu\,\,K_0(r) \,  , \label{eq:2B-wf-medium}
\end{eqnarray}
with $\lambda$ and $\mu$ being proportionality constants,
and $G_0$ and $K_0$ the evaluation of the vacuum and in-medium propagators
for zero external momenta:
\begin{eqnarray}
  G_0(r) &=& - m_N\int \frac{d^3 \vec{k}\,}{(2\pi)^3}\,\frac{e^{i \vec{k} \cdot \vec{r}}}{k^2} \nonumber \\
  &=& -\frac{m_N}{4 \pi}\,\frac{1}{r} \, , \\
  K_0(r) &=& - m_N\int \frac{d^3 \vec{k}\,}{(2\pi)^3}\,\frac{e^{i \vec{k} \cdot \vec{r}}}{k^2}\,\theta(k-k_F) \nonumber \\
  &=& -\frac{m_N}{4 \pi}\,\left[ \frac{1}{r} - \frac{2}{\pi}\,\frac{{\rm Si}(k_F r)}{r}\right] \, ,
\end{eqnarray}
where ${\rm Si}(x)$ is the sine-integral function:
\begin{eqnarray}
  {\rm Si}(x) = \int_0^x \frac{\sin{t}}{t}\, dt \, .
\end{eqnarray}
The proportionality constants can be determined from a comparison with
the equations
\begin{eqnarray}
  | \Psi_{\vec{p}\,} \rangle &=& | \vec{p}\, \rangle + \frac{1}{E-H_0} T(E)\,|\vec{p}\,\rangle   \, , \\
  | \Psi_{\vec{p}\, , \vec{P}\,} \rangle &=& | \vec{p}\, \rangle + \frac{Q_F}{E-H_0} G(E)\,|\vec{p}\,\rangle   \, , \label{eq:Psi-G-matrix}
\end{eqnarray}
where $T(E)$ and $G(E)$ are the T- and G-matrices, and with $Q_F$ the
projector that removes intermediate states within the Fermi sea.
After Fourier-transforming into r-space and taking the limit of zero external
momenta, it is apparent that $\lambda$ and $\mu$ can be identified with
$T(E)$ and $G(E)$ at $E=0$, respectively, where
\begin{eqnarray}
  T(E \to 0) &\to& \frac{4\pi}{m_N}\,a_0 \, , \\
  G(E \to 0) &\to& \frac{4\pi}{m_N}\,\frac{1}{\frac{1}{a_0} - \frac{2}{\pi}\,k_F}
  \, ,
\end{eqnarray}
with $E = p^2 / m_N$ the center-of-mass energy of the two-body system.
For the G-matrix the $P \to 0$ limit has been implicitly understood,
where a more detailed derivation of this limit can be
found in Appendix \ref{app:g-matrix}.

\begin{figure}[ttt]
  \begin{center}
    \includegraphics[width=7.5cm]{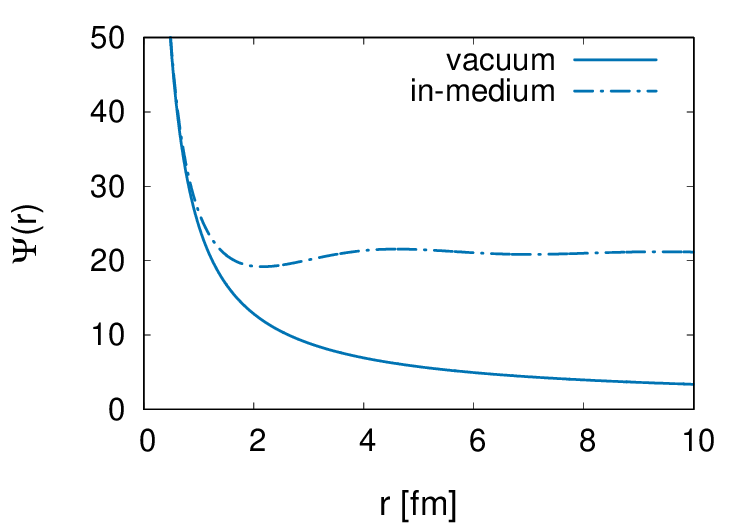}
\end{center}
  \caption{Vacuum and in-medium zero-energy, S-wave two-body wave functions.
    The normalization has been set from the condition that both wave functions
    coincide in the $r \to 0$ limit, where $\Psi(r) \to (1 - {a_0}/r)$
    with $a_0$ the scattering length.
    With this normalization in the asymptotic $r \to \infty$ limit the wave
    function approaches $\Psi \to 1$ in vacuum and
    $\Psi \to (1 - 2 a_0 k_F / \pi)$ in medium.
    The scattering length and Fermi momentum have been taken to be
    $a_0 = -23.7\,{\rm fm}$ and $k_F = 1.33\,{\rm fm}^{-1}$,
    which correspond to the singlet neutron-proton scattering length and
    the saturation density of symmetric nuclear matter.
}
\label{fig:in-medium-wf}
\end{figure}

Putting the pieces together,
the zero-energy two-body wave function in the medium reads
\begin{eqnarray}
  \Psi_{0,0}(\vec{r}\,) = 1 -
  \frac{1}{\frac{1}{a_0} - \frac{2}{\pi}\,k_F}\,
  \left[ \frac{1}{r} - \frac{2}{\pi}\,\frac{{\rm Si}(k_F r)}{r}\right] \, .
\end{eqnarray}
The vacuum solution is reproduced as the $k_F \to 0$ limit of
the in-medium wave function.
In Fig.~\ref{fig:in-medium-wf} a comparison of the vacuum and
in-medium solutions for $a_0 = -23.7\,{\rm fm}$
--- corresponding to the ${}^1S_0$ scattering length in the
two-nucleon system -- and $k_F = 1.33\,{\rm fm}$ (i.e. the Fermi momentum
of the saturation density of nuclear matter) is shown.
There one can clearly see how the in-medium wave function
becomes approximately constant for $k_F r \geq 1$, modulo
decreasing oscillations.

The oscillations are a well-known part of the asymptotic ($k_F r \gg 1$)
behavior of the wave function:
\begin{eqnarray}
  \Psi_{0,0}(\vec{r}\,) &\to& 1 - 
  \frac{\frac{2}{\pi} k_F}{\frac{1}{a_0} - \frac{2}{\pi}\,k_F}\,
  \left[ \frac{\cos{(k_F r)}}{{(k_F r)}^2} + \frac{\sin{(k_F r)}}{{(k_F r)}^3}
    \right] \nonumber \\
  && \quad + \mathcal{O}\left( \frac{1}{r^4} \right) \, ,
\end{eqnarray}
where it can be appreciated the pattern of $1/r^2$ and $1/r^3$
corrections over the free solution~\cite{Day:1967zza}.
It is interesting to notice the absence of a $1/r$ component, which is
related to scattering. This reflects the well-known fact that
the in-medium phase shifts are zero.
These oscillations are faintly visible in Fig.~\ref{fig:in-medium-wf}.

\subsection{In-medium renormalization group flow}

As is well known, the form of the two-body wave function contains in itself
interesting information about the renormalization of the two-body system
at low energies.
In particular, by introducing a cutoff it will be possible to analyze the RG
evolution of the contact-range couplings, determine their sizes
in the infrared limit and hence their power counting.
Of course, this is also the case at finite densities.

One might include a cutoff $R_c$ by regularizing the contact-range potential
as a delta-shell at the radius $r = R_c$:
\begin{eqnarray}
  V_C(r; R_c) = C_0(R_c)\,\frac{\delta(r-R_c)}{4\pi R_c^2} \, ,
\end{eqnarray}
with $C_0(R_c)$ the coupling constant.
Provided the wave function is known, the coupling and its running
are given by 
\begin{eqnarray}
  C_0(R_c) = \lim_{\epsilon \to 0^+}\frac{4\pi R_c^2}{m_N}\,
  \left( \frac{u_0'(R_c+\epsilon)}{u_0(R_c+\epsilon)} -
  \frac{u_0'(R_c-\epsilon)}{u_0(R_c-\epsilon)} \right) \, , \nonumber \\
  \label{eq:C0-running-delta-shell}
\end{eqnarray}
where $u_0(r) = r \, \Psi_0(r)$ is the zero-energy
reduced two-body wave function.

But first one needs the two-body wave functions for a delta-shell potential.
They can be obtained by solving the r-space Lippmann-Schwinger
equation and its in-medium counterpart, leading to
\begin{eqnarray}
  \Psi_0(r) &=& 1 + \lambda\,G_0(r; R_c) \,  , \\
  \Psi_{0,0}(r) &=& 1 + \mu\,\,K_0(r; R_c) \,  ,   
\end{eqnarray}
with $\lambda$, $\mu$ the same parameters that already
appeared in Eqs.~(\ref{eq:2B-wf-vacuum})
and (\ref{eq:2B-wf-medium}), and where the regularized
two-body propagators at zero external momenta are given by
\begin{eqnarray}
  && G_0(r; R_c)
  = \int_0^{\infty} dx\,\delta(x-R_c)\,\int \frac{d^2 \hat{x}}{4 \pi}
  G_0(|\vec{r} - \vec{x}\,|) \nonumber \\
  && \quad = - \frac{m_N}{4 \pi}\,\left[
    \frac{\theta(r-R_c)}{r} + \frac{\theta(R_c-r)}{R_c}
    \right] \, , \\
  && K_0(r; R_c) = \int_0^{\infty} dx\,\delta(x-R_c)\,\int \frac{d^2 \hat{x}}{4 \pi}
  K_0(|\vec{r} - \vec{x}\,|) \nonumber \\
    && \quad = - \frac{m_N}{4 \pi}\,\Bigg[
    \frac{\theta(r-R_c)}{r} + \frac{\theta(R_c-r)}{R_c} \nonumber \\
    && \qquad
    - \frac{2}{\pi r}\,\frac{f(k_F (r + R_c)) - f(k_F(r-R_c))}{2 k_F R_c} 
     \Bigg] \, , \label{eq:K0-delta-shell}
\end{eqnarray}
where the function $f$ is defined as
\begin{eqnarray}
  f(x) = - 1 + \cos{x} + x\, {\rm Si}(x) \, .
\end{eqnarray}

Thus when a cutoff is introduced the vacuum
zero energy wave function changes to
\begin{eqnarray}
    \Psi_{0}({r}) &=& 1 -
  a_0
  \Big[ \frac{\theta(r-R_c)}{r} + \frac{\theta(R_c-r)}{R_c} \Big]
  \, , \nonumber \\
\end{eqnarray}
in which it can be appreciated that the effect of the delta-shell is local:
for $r < R_c$ one ends up with a free wave function, while only
for $r > R_c$ are the effects of the regularized
contact-range interaction felt.
The corresponding in-medium wave function now changes to
\begin{eqnarray}
    \Psi_{0,0}({r}) &=& 1 -
  \frac{1}{\frac{1}{a_0} - \frac{2}{\pi}\,k_F}\,
  \Big[ \frac{\theta(r-R_c)}{r} + \frac{\theta(R_c-r)}{R_c} \nonumber \\
    && 
    - \frac{2}{\pi r}\,\frac{f(k_F (r + R_c)) - f(k_F(r-R_c))}{2 k_F R_c}\Big]
  \, , \nonumber \\ \label{eq:wf-in-medium-delta-shell}
\end{eqnarray}
where the second, $k_F$-dependent term in the parentheses is non-local.
That is, the effects of the delta-shell are now felt for all $r$.

\begin{figure}[ttt]
  \begin{center}
    \includegraphics[width=7.5cm]{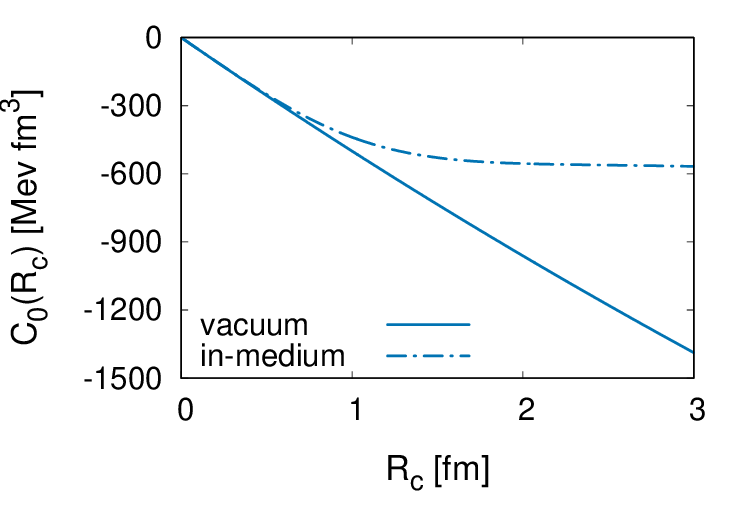}
\end{center}
  \caption{Running of the lowest order coupling constant $C_0$ with respect
    to the r-space cutoff $R_c$ in vacuum and in medium.
    The scattering length and Fermi momentum are the same
    as in Fig.~\ref{fig:in-medium-wf}.
}
\label{fig:in-medium-running}
\end{figure}

With the previous wave functions, one ends up with the couplings
\begin{eqnarray}
  \frac{1}{C_0(R_c)} &=& \frac{m_N}{4 \pi}\,\left( \frac{1}{a_0} - \frac{1}{R_c} \right) \, , \\
  \frac{1}{C_{0F}(R_c)} &=& \frac{m_N}{4\pi}\,\left(
  \frac{1}{a_0} - \frac{2}{\pi}\,k_F 
  - \frac{g_0(k_F R_c)}{R_c}
  \right) \, , \label{eq:C0F-delta-shell}
\end{eqnarray}
for the vacuum and in-medium case, respectively, where
\begin{eqnarray}
  g_0(x) = 1 - \frac{1}{\pi}\,\frac{1}{x}\,
  \left( -1 + \cos{(2x)} + 2x\,{\rm Si}(2x)\right) \, .
\end{eqnarray}
Their runnings are computed in Fig.~\ref{fig:in-medium-running}
for the specific case of $a_0 = -23.7\,{\rm fm}$ and $k_F = 1.33\,{\rm fm}^{-1}$,
corresponding to singlet neutron-proton scattering and
the saturation density of symmetric nuclear matter.
As can be readily noticed, the in-medium coupling RG evolution {\it freezes}
at distances $R_c \geq (1-2)\,{\rm fm}$, in agreement with
the naive intuition about its expected behavior.

It is also interesting to notice that the running of the in-medium coupling
implicitly contains resummations of $a_0 k_F$, which could be considered
a non-perturbative effect. Yet, here perturbativeness will be used
in the sense of expanding in terms of powers of the couplings,
regardless of whether the value of the couplings is derived
from a resummation in a different power counting.

\subsection{Infrared RG flow and power counting}

The power counting of the lowest-order contact-range theory is easy to
understand by evolving the couplings towards the infrared,
defined as $Q R_c \to 1$ (i.e. $R_c \to \infty$ for $Q \to 0$), and
then considering iterations of the contact within this limit.

For the delta-shell regulator the $R_c \to \infty$ limit of the
running of the vacuum coupling is
\begin{eqnarray}
  \frac{1}{C_0(R_c \to \infty)} &\to& \frac{m_N}{4 \pi}\,\frac{1}{a_0} \, .
  \label{eq:coupling-infrared-vacuum}
\end{eqnarray}
Thus, depending on the size of the scattering length,
this leads to the well-known scalings~\cite{vanKolck:1998bw,Chen:1999tn}
\begin{eqnarray}
  C_0(R_c \sim 1/Q) &\propto& \frac{4 \pi}{m_N}\,\frac{1}{M} 
  \quad \mbox{for $a_0 \sim 1/M$} \, , \\
  C_0(R_c \sim 1/Q) &\propto& \frac{4 \pi}{m_N}\,\frac{1}{Q} 
  \quad\,\, \mbox{for $a_0 \sim 1/Q$} \, .
\end{eqnarray}
For the in-medium coupling one has instead
\begin{eqnarray}
  \frac{1}{C_{0F}(R_c \to \infty)} 
  &\to& \frac{m_N}{4\pi}\,\left(
  \frac{1}{a_0} - \frac{2}{\pi}\,k_F \right) \, ,
  \label{eq:coupling-infrared-medium}
\end{eqnarray}
yielding the infrared scalings 
\begin{eqnarray}
  C_{0F}(R_c \sim 1/Q) &\propto&
  \frac{4 \pi}{m_N}\,\frac{1}{M} 
  \quad \mbox{for $\frac{2}{\pi}\,k_F a_0 \ll 1$} \, , \\
  C_{0F}(R_c \sim 1/Q) &\propto& \frac{4 \pi}{m_N}\,\frac{1}{Q} 
  \quad\,\, \mbox{for $\frac{2}{\pi}\,k_F a_0 \gg 1$} \, ,
\end{eqnarray}
which are like the vacuum ones but modified by
the presence of the Fermi momentum.

If one considers the iterations of the contact-range potential,
their sizes are easily determined from the perturbative expansion of
the Lippmann-Schwinger (Bethe-Goldstone) equation
for the T- (G-) matrix.
In the vacuum the expansion of the T-matrix is given by
\begin{eqnarray}
  T &=& \sum_{n \geq 1} T_n 
  = C_0(R_c) + G_0(R_c; R_c)\,C_0^2(R_c) + \dots \nonumber \\
  &=& C_0(R_c)- \frac{m_N}{4 \pi}\,C_0^2(R_c)\,\frac{1}{R_c} + \dots \, ,
\end{eqnarray}
where the subindex in $T_n$ refers to the number of iterations of
the contact-range coupling.
From this, in the infrared limit the iterations scale as
\begin{eqnarray}
  -\frac{T_{n+1}}{T_{n}} &=& \frac{m_N}{4 \pi}\,C_0(R_c)\,\frac{1}{R_c}
  \nonumber \\
  &\to& \frac{a_0}{R_c} \propto Q a_0 \, ,
\end{eqnarray}
where the second line corresponds to the $R_c \to \infty$ limit.
This implies a perturbative (non-perturbative) EFT if low-energy
scattering is weak (strong).

If one now considers the in-medium case, the expansion of
the G-matrix is given by
\begin{eqnarray}
  G &=& \sum_{n \geq 1} G_n 
  = C_{0F}(R_c) + K_0(R_c; R_c)\,C_{0F}^2(R_c) + \dots \nonumber \\
  &=& C_{0F}(R_c) - \frac{m_N}{4 \pi}\,C_{0F}^2(R_c)\,
  \frac{g_0(k_F R_c)}{R_c} + \dots \, ,
\end{eqnarray}
and thus in the infrared limit one has
\begin{eqnarray}
  -\frac{G_{n+1}}{G_{n}} &=& \frac{m_N}{4 \pi}\,C_{0F}(R_c) \frac{g_0(k_F R_c)}{R_c}
  \nonumber \\
  &\to& +\frac{1}{2}\,
  \frac{\frac{2}{\pi}\,k_F}{\frac{1}{a_0} -\frac{2}{\pi}\,k_F}\,
  \frac{1}{(k_F R_c)^2} \, ,
\end{eqnarray}
which implies that loops are to be counted as
\begin{eqnarray}
  \frac{G_{n+1}}{G_{n}} \propto \frac{Q}{M}\,\frac{Q}{k_F} \quad \mbox{or} \quad
          {\left( \frac{Q}{k_F} \right)}^2
\end{eqnarray}
for $a_0 \sim 1/M$ and $a_0 \sim 1/Q$, respectively.

This argument suggests that in nuclear matter (where the scattering length
is large) the lowest order contact is to be iterated at all orders,
which relates to previous discussions
in the literature~\cite{Yang:2016nkd,Grasso:2017jmr,Lacroix:2017whm}.
Yet, the infrared running of the loops is clearly different
than in the vacuum, leaving room for their numerical
suppression for values of $k_F R_c$ still of order
unity but somewhat larger than one.

If anything this asks for a more detailed examination of
the behavior of the loops, particularly owing to
the fact that the previous argument relies on
their bare (i.e. cutoff dependent) evaluation.
Thus, it can be improved by considering the renormalized loops instead,
for which one includes corrections to the coupling
constant that remove the cutoff dependence of the loops.
For this, one expands the coupling around the infrared limit,
which for the vacuum case yields
\begin{eqnarray}
  C_0(R_c) &=&
  \frac{4 \pi}{m_N}\,a_0 + \frac{4 \pi}{m_N}\,\frac{a_0^2}{R_c} + \dots
  \nonumber \\
  &=& C_0 + \delta C_0 + \dots \, ,
\end{eqnarray}
where in the present context $C_0$ refers to the finite part of $C_0(R_c)$
in the $R_c \to \infty$ limit and $\delta C_0$ to its first,
cutoff-dependent correction. 
Now the perturbative series for the T-matrix reads
\begin{eqnarray}
  \sum_{n=0}^{\infty} T_n = C_0 + \delta C_0 -
  \frac{m_N}{4 \pi}\,\frac{C_0^2}{R_c}  + \dots \, ,
\end{eqnarray}
which leads to the conclusion that there are no perturbative
corrections to the T-matrix at zero energy ($E = 0$),
or equivalently, at zero on-shell momentum
($p = \sqrt{m_N E} = 0$):
\begin{eqnarray}
  T(p=0) = \frac{4\pi}{m_N}\,a_0 = C_0 \, .
\end{eqnarray}
In hindsight this is trivial: at zero energy one can always rearrange
the higher order corrections to the coupling in such a way
that the T-matrix does not change.

Instead, what one has to consider is the expansion in terms of
powers of the external momentum $p$, whose effect is to
change the propagator to
\begin{eqnarray}
  G_0(R_c; R_c) = - \frac{m_N}{4 \pi}\,\left[ \frac{1}{R_c} + i\,{p} +
    \mathcal{O}(p^2) \right] \, ,
\end{eqnarray}
in which case one ends up with
\begin{eqnarray}
 && T(p) = C_0 + \delta C_0 -
       \frac{m_N}{4 \pi}\,C_0^2\,
       \left[ \frac{1}{R_c} + i\,p \right]
       + \dots \nonumber \\
 && \qquad = \frac{4 \pi}{m_N}\,a_0\,\left( 1 - i\,a_0 p + \dots \right) \, ,
\end{eqnarray}
which recovers the first non-trivial term in the expansion of the T-matrix
for a zero-range interaction, that is:
\begin{eqnarray}
  T(p) = \frac{4\pi}{m_N}\,\frac{1}{\frac{1}{a_0} + i\,p} =
  \frac{4 \pi}{m_N}\,a_0\,\sum_{n=0}^{\infty}\,{( - i a_0 p )}^n \, .
\end{eqnarray}
In turn, this also implies the well-known result that loops scale as $Q a_0$.

For the iteration of the in-medium coupling, one begins with the infrared
expansion of the coupling itself
\begin{eqnarray}
  C_{0F}(R_c) &=&
  \frac{4 \pi}{m_N}\,\frac{1}{\frac{1}{a_0} - \frac{2}{\pi}\,k_F}
  \nonumber \\
  &+& \frac{4 \pi}{m_N}\,
  \frac{1}{{\left( \frac{1}{a_0} - \frac{2}{\pi}\,k_F \right)}^2}\,
  \frac{g_0(k_F R_c)}{R_c} + \dots
  \nonumber \\
  &=& C_{0F} + \delta C_{0F} + \dots \, ,
  \label{eq:C0F-IR-expansion}
\end{eqnarray}
and then considers the expansion of the loops
in powers of the external total and relative momenta
\begin{eqnarray}
  K_0(R_c;R_c) &=& -\frac{m_N}{4 \pi}\,\Bigg[
    \frac{g_0(k_F R_c)}{R_c} \nonumber \\ &+&
    P\,g_{1}(k_F R_c) 
    + \mathcal{O}\left( p^2, P^2 \right) \Bigg] \, ,
\end{eqnarray}
where the term proportional to $p$ vanishes (as scattering
is suppressed below the Fermi sea).
Here $g_{1}$ is given by
\begin{eqnarray}
  g_{1}(x) &=& - \frac{1}{\pi}\,\frac{1- \cos{(2x)}}{(2x)^2} \, .
\end{eqnarray}
Thus, up to corrections proportional to $P$,
the expansion of the G-matrix reads
\begin{eqnarray}
  G(p,P) &=& G + \delta G + \dots \nonumber \\
  &=& C_{0F} + \delta C_{0F} \nonumber \\
  &-& \frac{m_N}{4 \pi} \,C_{0F}^2\, 
       \Big[ \frac{g_0(k_F R_c)}{R_c} + P\,g_{1}(k_F R_c) + \dots \Big]
   \nonumber \\  &+& \dots \, ,
\end{eqnarray}
where $C_{0F}$ and $\delta C_{0F}$ (without reference to a specific cutoff)
refer to the expansion of $C_{0F}(R_c)$ in the infrared,
Eq.~(\ref{eq:C0F-IR-expansion}).
After removing the divergence at $p=P=0$ (i.e., the cancellation of
the $1/R_c$ cutoff dependence term that happens when adding
the $\delta C_{0F}$ and $C_{0F}^2$ contributions),
the relative size of the corrections is
\begin{eqnarray}
  -\frac{\delta G}{G} &=&
   \frac{m_N}{4 \pi} C_{0F}\,
   P\,g_1(k_F R_c) \nonumber \\
  &=& 
  \frac{1}{\frac{1}{a_0} - \frac{2}{\pi}\,k_F}\,
  \frac{P}{\pi}\,\frac{\cos{(2 k_F R_c)} - 1}{(2 k_F R_c)^2} \nonumber \\
  &\propto& \left( \frac{Q}{k_F} \right)\,\frac{\sin^2{(k_F R_c)}}{2 (k_F R_c)^2}
  \, ,
\end{eqnarray}
where the second line assumes $\frac{2}{\pi}\,a_0 k_F \gg 1$ (e.g. unnatural
scattering length and nuclear matter equilibrium density).

This reproduces the previous estimate about the infrared scaling of the loops,
but adds a new twist: the loop contributions might very well vanish at
certain cutoffs.
For the case at hand, the first r-space cutoff at which this particular
loop contribution vanishes is $k_F R_c = \pi$.

In general, if one expands the in-medium loop function further in powers of
$P$ and $p$, then the individual terms in this expansion do not necessarily
vanish at the same cutoffs (and some of them do not vanish at any cutoff).
The proper comparison to be made is not the perturbative expansion of
the G-matrix, but rather the perturbative expansion of the potential
energy per nucleon (or per fermion).

This is nicely illustrated by considering the next two terms in the expansion
of the propagator as a function of the external momenta, i.e. the terms
proportional to $p^2$ and $P^2$:
\begin{eqnarray}
  K_0(R_c;R_c) &=& -\frac{m_N}{4 \pi}\,\Bigg[
    \frac{g_0(k_F R_c)}{R_c} 
    + P\,g_{1}(k_F R_c)  \nonumber \\ 
 &+&  \frac{p^2}{k_F}\,g_{2A}(k_F R_c) + \frac{P^2}{k_F}\,
    g_{2B}(k_F R_c) \nonumber \\
    &+& \mathcal{O}\left( p^3, P^3, P p^2 \right)
    \Bigg] \, , 
\end{eqnarray}
with
\begin{eqnarray}
  g_{2A}(x) &=& - \frac{2}{3}\,x + \frac{1}{3\pi\, x^2} + \frac{(2 x^2- 1)\,\cos{(2x)}}{3\pi\, x^2} \nonumber \\
  &+& \frac{\sin{(2 x)}}{3\pi\, x} + \frac{4}{3\pi}\,x\,{\rm Si}(2 x)
  \, , \\
  g_{2B}(x) &=& \frac{(2\sin(x) - x\,\cos(x))\,\sin{x}}{6 \pi x^2} \, .
\end{eqnarray}
In this case the relative size of the renormalized loop corrections is
\begin{eqnarray}
 - \frac{\delta G}{G} &=&
 \frac{m_N}{4 \pi} C_{0F}\,
 \Bigg[ P\,g_1(k_F R_c) \nonumber \\ && \quad +
   \frac{p^2\,g_{2A}(k_F R_c) + P^2\,g_{2B}(k_F R_c)}{k_F}
   \nonumber \\ && \quad + \mathcal{O}\left( p^3, P^3, P p^2 \right)
   \Bigg] \, ,
\end{eqnarray}
where $g_1$ and $g_{2B}$ do both have zeros, though not at the same
cutoff, while $g_{2A}$ is always positive.
This indicates that power counting arguments solely grounded in the loop
corrections to the G-matrix are not necessarily straightforward,
as there is a certain cutoff asynchrony among the different
terms in its low-momentum expansion.

This ceases to be a problem if one considers instead the potential energy
contribution to the EOS (as in~\cite{Furnstahl:1999ix,Steele:2000qt}),
i.e., the average of the G-matrix in the medium,
which is proportional to
\begin{eqnarray}
  && (E + \delta E) \propto \nonumber \\ && \quad
  \Omega\,\int_{k_1, k_2 < k_F}
  \frac{d^3\,\vec{k}_1}{(2 \pi)^3}\,\frac{d^3\,\vec{k}_2}{(2 \pi)^3}\,
  \left( G + \delta G \right) \, ,
\end{eqnarray}
where $\Omega$ is the volume of the system. In this case one arrives at
\begin{eqnarray}
  && - \frac{\delta E}{E} \propto \frac{m_N k_F}{4 \pi}\,C_{0F}\,\Bigg[
    \frac{36}{35}\,g_1(k_F R_c) \nonumber \\ && \quad
    + \frac{3}{10}\,g_{2A}(k_F R_c) + \frac{6}{5}\,
    g_{2B}(k_F R_c) + \dots \Bigg] \, ,
\end{eqnarray}
where the sum of the terms within the brackets do now have zeros
in the infrared limit.
For the three terms included in the brackets, the first zero happens
at $k_F R_c \approx 0.8498$ (compatible with $Q R_c \sim 1$, i.e.,
with the infrared limit).

If one were to include all the powers in the expansion
on the external momenta of $G(p,P)$ and $\delta G(p,P)$, one could simply write
\begin{eqnarray}
  -\frac{\delta E}{E} &\propto& \frac{m_N k_F}{4 \pi}\,C_{0F}\,h_{F}(k_F R_c) \, ,
  \label{eq:hF-ratio}
\end{eqnarray}
where the $h_F(x)$ function represents the relative strength of
the loop contributions to the potential energy
in nuclear matter.
If scattering is strong at low energies, as is the case
in the two-nucleon system, the expression simplifies to
\begin{eqnarray}
  \frac{\delta E}{E} &\propto& \frac{\pi}{2}\,h_{F}(k_F R_c) \, ,
\end{eqnarray}
which further cements the interpretation of $h_F$ as
the effective expansion parameter for the loops.

\begin{figure}[ttt]
  \begin{center}
    \includegraphics[width=7.5cm]{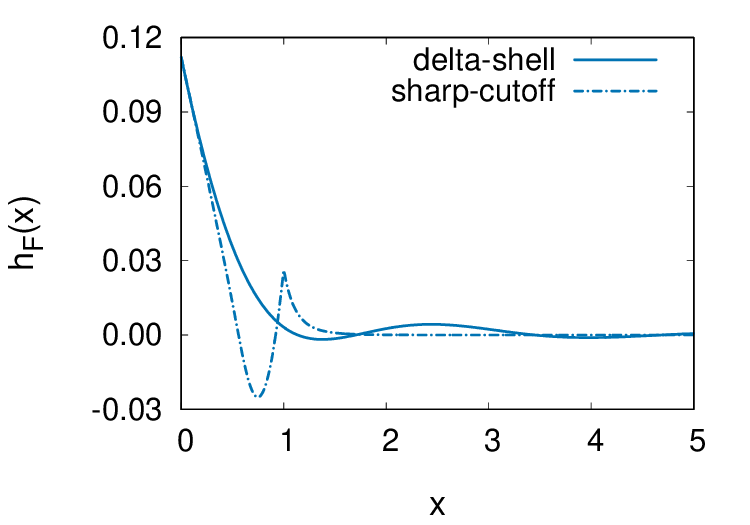}
\end{center}
  \caption{The function $h_F(x)$, which determines the relative size of
    the loops in nuclear matter, as calculated for a delta-shell
    regulator in r-space and a sharp-cutoff in p-space.
    Here $x = k_F R_c$ or $x = k_F / \Lambda$ for the delta-shell and
    sharp-cutoff, respectively, with $R_c$ and $\Lambda$ the r-space
    and p-space cutoffs.
}
\label{fig:in-medium-loop}
\end{figure}

For the delta-shell regulator there is no analytical expression
for $h_{F}(x)$, which has to be evaluated numerically instead.
This is done in Fig.~\ref{fig:in-medium-loop}, where it can be appreciated
that regardless of the cutoff the loops are numerically suppressed.
The first zero of the function $h_F(x)$ happens at $x \approx 1.137$,
followed by $x \approx 1.701$, then $x \approx 3.451$ and so on,
which for $k_F = 1.33\,{\rm fm}^{-1}$ (the equilibrium density of
nuclear matter) translates into $R_c = 0.85, 1.28, 2.59\,{\rm fm}$.
Regardless of whether the loop contribution vanishes for a given cutoff,
the fact is that $h_F(x)$ becomes rather small once the infrared
regime ($x \geq 1$) is reached.
This reinforces the observation that the loops are indeed
suppressed in the infrared, hence implying a perturbative
power counting for nuclear matter.

For assessing the regulator dependence of the previous result one might
consider a sharp cutoff regulator in momentum space,
for which the running of the in-medium coupling is
\begin{eqnarray}
  C_{0F}(\Lambda) = C_0(\Lambda)\,\theta(\Lambda - k_F) +
  C_0(k_F)\,\theta(k_F - \Lambda) \, , \nonumber \\
  \label{eq:C0F-sharp-cutoff}
\end{eqnarray}
with $C_0(\Lambda)$ the vacuum running
\begin{eqnarray}
  \frac{1}{C_0(\Lambda)} =
  \frac{m_N}{4 \pi}\,\left( \frac{1}{a_0} - \frac{2}{\pi}\,\Lambda \right) \, .
\end{eqnarray}
That is, for a sharp cutoff the $C_0$ coupling {\it exactly} freezes
for $\Lambda < k_F$, providing the ideal illustration of this phenomenon.
The calculation of $h_F(x)$ can be done analytically in this case, though
the expression is rather cumbersome (it can be found
in Eq.~(\ref{eq:hF-SC-I}) of Appendix \ref{app:g-matrix}
for $x \leq 1/2$).
Its numerical evaluation is shown in Fig.~\ref{fig:in-medium-loop},
where there are now only two zeros located at $x \approx 0.5592$ and
$x \approx 0.9210$, with $x = k_F / \Lambda$.
Even though the specific details of the infrared behavior are different
from those of the delta-shell (for instance, the derivative of $h_F(x)$ shows
a discontinuity at $x=1$ corresponding to the one in the running of the
in-medium coupling), the loops are nonetheless still suppressed.

It is also worth noticing that even in the ultraviolet limit ($x \to 0$)
one finds that
\begin{eqnarray}
  h_F(0) = \frac{2}{\pi} - \frac{6\,(11-2\log{2})}{35\pi} \approx 0.11203 \, ,
  \label{eq:uv-in-medium-loop}
\end{eqnarray}
which is numerically small and regulator independent, where the second
term is very familiar as it appears frequently in calculations
of many-fermion systems (e.g. in~\cite{Lee:1957zza}
from the fifties).
This limit is not entirely a curiosity in the sense that
this also corresponds to the calculation of $h_F(x)$
in power divergence subtraction (PDS)~\cite{Kaplan:1998tg,Kaplan:1998we},
a modification of dimensional regularization that is particularly
well suited for the two-nucleon system~\footnote{
  A quirk of PDS is that the running of the vacuum and in-medium
  couplings is identical: for a contact-range theory,
  PDS is only able to see the linear divergence of
  the loop integrals at the expense of
  other types of cutoff dependence.
  That is, the healing distance sits squarely
  within the blind spot of PDS.
  Yet, there is a simple trick for being able to recycle the previous formalism
  with PDS, which is to reorder the coupling as
  \begin{eqnarray}
    C_{0F}^{\rm PDS}(\Lambda) &=& \frac{4\pi}{m_N}\,
    \frac{1}{\frac{1}{a_0} - \frac{2}{\pi}\,k_F - (\Lambda - \frac{2}{\pi} k_F)}
    \, , \nonumber
  \end{eqnarray}
  and expand in powers of $(\Lambda - \frac{2}{\pi} k_F)$.
}.

In terms of power counting the conclusion is that in-medium loops are suppressed
in the infrared limit with respect to their behavior in the vacuum.
This suppression is not the usual one involving the $Q / M$ ratio of
the characteristic soft and hard scales of the theory,
but a numerical one instead.
Of course a warning is in place: this conclusion is dependent on relying
on the in-medium running of the coupling and also on the choice of
regulator, on which the numerical suppression depends.
Indeed the arguments presented here require the use of the in-medium running.
This often assumes full knowledge of the in-medium two-body wave function,
which in general is not the case, precluding the exact calculation
of the infrared limit of the in-medium running of the coupling.
Of course there will be exceptions and certain regulators might provide
remarkable simplifications.

With this limitation in mind, one might consider using the infrared
limit of the vacuum coupling $C_0$ instead,
Eq.~(\ref{eq:coupling-infrared-vacuum}).
By repeating the previous arguments, one finds:
\begin{eqnarray}
  - \frac{\delta E}{E} \propto a_0 k_F\,
  \left[ h_F(k_F R_c) - \frac{2}{\pi} \right] \, ,
\end{eqnarray}
which results in the subtraction of a factor of $2/\pi$ to the dimensionless
loop function $h_F(x)$.
This factor originates from the difference between the infrared limits of
$C_0(R_c)$ and $C_{0F}(R_c)$ and its interpretation is straightforward:
the loop corrections are now forced to recover this difference,
which is not perturbative (in the sense that $C_{0F}$ contains
resummations of $k_F a_0$ in the infrared).

Yet, there is a simple workaround grounded in a trivial observation:
the value of the vacuum coupling at a finite cutoff $R_c^*$ is
basically equivalent to the infrared limit of the in-medium
coupling at some Fermi momentum $k_F^*$.
For a contact-range theory and a delta-shell regulator
the equivalence in the $R_c^* \to \infty$ limit
is given by $k_F^* R_c^* = \pi/2$.
Outside this limit, one might recalculate the renormalized loop correction
\begin{eqnarray}
  \frac{\delta E}{E} \propto -\frac{m_N}{4 \pi}\,C_0(R_c^*)\,
  \left[ \frac{1}{R_c^*} + k_F \left( h_F(k_F R_c) - \frac{2}{\pi} \right)
    \right] \, ,
  \nonumber \\
\end{eqnarray}
then define the function
\begin{eqnarray}
  \bar{h}_F(x) = h_F(x) - \frac{2}{\pi} \, ,
    \label{eq:uv-in-medium-loop-modified}
\end{eqnarray}
and from it, the equivalent Fermi momentum
\begin{eqnarray}
  1 + (k_F^* R_c^*)\,\bar{h}_F (k_F^* R_c^*) = 0 \, ,
\end{eqnarray}
leading to
\begin{eqnarray}
  \frac{\delta E}{E} \propto \frac{k_F^* \bar{h}_F(k_F^* R_c^*) -
    k_F \bar{h}_F(k_F R_c)}{k_F^* \bar{h}_F(k_F^* R_c^*)} \, .
\end{eqnarray}
This expression is not transparent until one considers the infrared limit,
for which $\bar{h}_F \to -2/\pi$ (i.e. $h_F \to 0$), in which case:
\begin{eqnarray}
  \frac{\delta E}{E} \propto \frac{k_F^* - k_F}{k_F^*} \, .
\end{eqnarray}
That is, even if one does not know the exact running of the in-medium coupling
(which explicitly depends on the Fermi momentum), it is still possible to use
the vacuum coupling instead and choose a cutoff $R_c^*$  such
that loop corrections cancel for a Fermi momentum $k_F^*$,
which depends on the previous cutoff.
This will be the approach followed in the next Section.

\section{Power counting in nuclear matter}
\label{sec:nuclear-matter}

Now I will consider nuclear matter, understood as a non-relativistic
system of fermions with two (spin) or four (spin and isospin)
degrees of freedom, where the first case corresponds to neutron matter and
the second to symmetric nuclear matter.
For simplicity, Wigner SU(4) will be assumed and only S-wave contact-range
interactions will be examined in detail. The focus is on power counting,
for which S-wave interactions probably provide the best example
of how it works, and where the differences between singlet and
triplet S-wave scattering are in principle not crucial.
P-wave and higher orbital angular momentum interactions usually enter
at subleading orders, and their power counting relies on the same
set of ideas as S-wave interactions.

First, I will deal with the power counting of two-body forces, which
illustrate the most important issues regarding
the organization of EFT in nuclear matter.
After that, I will briefly discuss three- and four-body forces.

\subsection{The two-body potential}

The form of the two-body contact-range potential depends on the explicit
representation chosen for it.
If one considers momentum-dependent interactions, it will be given by
\begin{eqnarray}
  V_{2B} = \sum_{n \geq 0}\, C_{2n}(\Lambda)\,\frac{1}{2}(p^{2n} + {p'}^{2n}) \, ,
  \label{eq:VC-momentum-dependent}
\end{eqnarray}
with $C_{2n}$ denoting the couplings and $p$, $p'$ the initial and final
relative momenta of the nucleons.
Alternatively, one may use the energy-dependent representation
\begin{eqnarray}
  V_{2B} = \sum_{n \geq 0}\, C_{2n}(\Lambda)\,k^{2n} \, ,
\end{eqnarray}
where $k = \sqrt{m_N E}$ with $m_N$ the nucleon mass and $E$
the center-of-mass energy of the two nucleons.
It happens that in nuclear matter it is possible to switch energy-dependence
with density-dependence, which leads to the following representation
\begin{eqnarray}
  V_{2B} = \sum_{n \geq 0}\, C_{2n}(\Lambda)\,k_F^{2n} \, ,
\end{eqnarray}
where $k_F$ is now the Fermi momentum.
This last representation is the most convenient by far for dealing with
nuclear matter, and is the one I will use as the default case.

The previous three representations are equivalent at tree level
in terms of the equation of state (EOS).
But the EOS is not an observable and this equivalence
breaks down once loop corrections are included.
Besides, while the momentum- and energy-dependent representation provide
a direct bridge with two-nucleon forces and the properties of
two-nucleon systems (e.g. phase shifts and deuteron
observables),
this is not the case for the density-dependent representation,
which can instead be thought as nuclear matter specific
(here one may check~\cite{Furnstahl:2000we} for an EFT-specific
discussion of off-shell effects at finite density).
It is also worth mentioning that density-dependent interactions induce
thermodynamic instabilities in the EOS, which can nonetheless be
corrected by the inclusion of additional terms~\cite{Hugenholtz:1958zz}.

\subsection{Equation of state at tree level}

At tree level, with the previous contact-range potential,
one can easily calculate the energy per nucleon as
\begin{eqnarray}
  \frac{E}{A} &=&
  \frac{\Omega}{A}\,\sum_{s_1}\,\int_{k_1 < k_F}
  \frac{d^3 \vec{k}_1}{(2\pi)^3} \frac{k_1^2}{2 m}
  \nonumber \\
  &+& \frac{\Omega}{2 A}\,\sum_{s_1,s_2}\,\int_{k_1, k_2 < k_F}
  \frac{d^3 \vec{k}_1}{(2\pi)^3}\,\frac{d^3 \vec{k}_2}{(2\pi)^3}
  \,\langle V_{2B} \rangle \, ,
  \nonumber \\
\end{eqnarray}
where $s_1, s_2$ are the spins (or isospins) of nucleons $1$, $2$, and
$\langle V_{2B} \rangle$ is the matrix element of $V_{2B}$ when sandwiched
between antisymmetric ${| \vec{k}_1 s_1 \vec{k}_2 s_2 \rangle}_A$ two-body
states.

If there are only S-wave two-body forces one arrives at the EOS
\begin{eqnarray}
  \frac{E}{A} = t \, \rho^{2/3} + c_0 \, \rho + c_2 \, \rho^{5/3} +
  \sum_{n \geq 2} c_{2n} \, \rho^{1 + 2 n / 3} \, ,
\end{eqnarray}
where $\rho$ is the density and $t$, $c_{2n}$ parameters for the kinetic and
potential energy terms.
The density is given by
\begin{eqnarray}
  \rho = \sum_{s_1}\,\int_{k_1 < k_F} \frac{d^3 \vec{k}_1}{{(2\pi)}^3} =
  g\,\frac{k_F^3}{6 \pi^2} \, , 
\end{eqnarray}
with $g$ the number of different spin states ($g=2$ for neutron matter
and $4$ for symmetric nuclear matter).
Meanwhile, $t$ is given by 
\begin{eqnarray}
  t &=& \frac{3}{5}\,{\left( \frac{6\pi^2}{g} \right)}^{2/3}\,\frac{1}{2 m_N}
  \, ,
\end{eqnarray}
while the $c_{2n}$ coefficients read (in the density-dependent representation)
as
\begin{eqnarray}
  c_{2n} &=& \frac{1}{2}(1-\frac{1}{g})\,C_{2n}(\Lambda)\,
  {\left( \frac{6\pi^2}{g} \right)}^{2n/3} \, .
\end{eqnarray}

\subsection{Loops and the divergence structure}

The one-loop correction to the tree level EOS can be written as
\begin{eqnarray}
  \frac{\delta E}{A} = \frac{\Omega}{2 A}\,\sum_{s_1, s_2}\,
  \int_{k_1, k_2 < k_F} \frac{d^3 \vec{k}_1}{(2\pi)^3} \frac{d^3 \vec{k}_2}{(2\pi)^3}\,
  \langle V_{2B} \frac{Q_F}{e_{12}} V_{2B} \rangle \, ,
  \nonumber \\
\end{eqnarray}
where the matrix element corresponding to the first iteration of the two-body
potential is
\begin{eqnarray}
  \langle V_{2B} \frac{Q_F}{e_{12}} V_{2B} \rangle =
  \int \frac{d^3 \vec{p}\,'}{(2\pi)^3} \,
  \frac{{| \langle \vec{p}\, | V_{2B} | \vec{p}\,' \rangle |}^2}{e_{12}}\,
  Q_{F} \, ,
\end{eqnarray}
where the spin indices have been omitted.
Here $e_{12}$ is the energy denominator 
\begin{eqnarray}
  e_{12} &=& \frac{{k}_1^2}{2m_N} + \frac{{k}_2^2}{2m_N} -
  \frac{{{k}_1'}^2}{2m_N} - \frac{{{k}_2'}^2}{2m_N} \nonumber \\
  &=& \frac{p^2}{m_N} - \frac{{p'}^2}{m_N} \, ,
\end{eqnarray}
which only depends on the initial and final relative momenta
$\vec{p}\,$ and $\vec{p}\,'$ (as the total momentum
is conserved in the loops), and $Q_{F}$ is the
projector into states over the Fermi momentum
\begin{eqnarray}
  Q_F  = \theta(k_1' - k_F)\,\theta(k_2' - k_F) \, ,
\end{eqnarray}
where
\begin{eqnarray}
  \vec{k}_1 = \frac{\vec{P}}{2} + {\vec{p}} \quad &\mbox{,}& \quad
  \vec{k}_2 = \frac{\vec{P}}{2} - {\vec{p}} \, , \nonumber \\
  \vec{k}_1\,' = \frac{\vec{P}\,'}{2} + {\vec{p}\,'} \quad &\mbox{and}& \quad
  \vec{k}_2\,' = \frac{\vec{P}\,'}{2} - {\vec{p}\,'} \, .
\end{eqnarray}
That is, $\vec{k}_j$, $\vec{k}_j'$ are the momenta of
the individual particles, and $\vec{P}$, $\vec{p}$ and
$\vec{P}\,'$, $\vec{p}\,'$ are the total and
center-of-mass momenta.
Owing to momentum conservation in the loops,
one has that $\vec{P} = \vec{P}\,'$.

The loop integral is divergent and requires regularization.
By simply cutting it at $p = \Lambda$ and focusing on the
linear divergence, one obtains
\begin{eqnarray}
  && \langle V_{2B} \frac{Q_F}{e_{12}} V_{2B} \rangle =
  -\frac{m_N}{2 \pi^2}\,\Lambda
  \nonumber \\
  && \times
  \left[ C_0^2 + 2\,C_0 C_2\,k_F^2 + (C_2^2 + 2 C_0 C_4)\, k_F^4 + \dots \right] 
  \nonumber \\
  && + \,\, \mbox{(finite terms)} \, ,
\end{eqnarray}
where it is now apparent why the density-dependent representation
has been chosen.
By integrating over the initial momenta (which are all below the Fermi momentum)
and rearranging the terms, one arrives at
\begin{eqnarray}
  \frac{E}{A} + \frac{\delta E}{A} &=& t\,\rho^{2/3} \nonumber \\
  &+&
  \left[ c_0 + \delta c_0 + \lambda_0 \,c_0^2\, \Lambda \right] \rho \nonumber \\
  &+& \left[ c_2 + \delta c_2 + 2\,\lambda_0 \,c_0 c_2\,\Lambda \right] \rho^{5/3}
  \nonumber \\
  &+& \left[ c_4 + \delta c_4 + \lambda_0\,(c_2^2 + 2 c_0 c_4)
    \Lambda \right] \rho^{7/3}
  \nonumber \\
  &+& \dots \, ,
\end{eqnarray}
where the dots indicate either finite terms or divergent terms involving
higher powers of the density, while $\lambda_0$ is a numerical constant.

From this it is apparent that the iteration of $C_0$ does not generate
new divergent terms depending on higher powers of the density and is
thus self-contained.
In the spirit of the arguments given in~\cite{Yang:2013bhd,Yang:2016mvq}
this defines a natural candidate for the lowest-order approximation,
one in which only $C_0$ is included, and for which the first
loop correction is given by
\begin{eqnarray}
  \frac{E}{A} + \frac{\delta E}{A} &=& t \rho^{2/3} + c_0 \rho \nonumber \\
  &+& \left[ \delta c_0 + \lambda_0 \, c_0^2 \Lambda \right] \rho +
  \dots \, ,
\end{eqnarray}
where the dots indicate finite contributions only (which for $C_0$
have a density dependence of $\rho^{4/3}$).
By decomposing the loop correction into a renormalized and divergent part
\begin{eqnarray}
  \delta c_0 = \delta c_0^R  - \lambda_0 \, c_0^2 \, \Lambda \, ,
\end{eqnarray}
one ends up with
\begin{eqnarray}
  \frac{E}{A} + \frac{\delta E}{A} &=& t\,\rho^{2/3} + (c_0 + \delta c_0^R)
  \rho + \dots \, , 
\end{eqnarray}
which basically only induces a finite correction to the $c_0$ parameter.
This correction could in principle be zero if the original $c_0$ parameter
has been optimally chosen in the first place~\footnote{This specifically
  applies to the correction of the $c_0 \rho$ term in the EOS. There is
  still a non-trivial, finite loop correction proportional to
  $c_0^2 \, \rho^{1+1/3}$, i.e. a term with a higher density
  dependence.}.

In contrast, the corrections to the $C_2$, $C_4$ and higher derivative
couplings are not self-contained and include divergent terms dependent
on higher powers of the density (this point is intimately
related to the discussion in~\cite{Yang:2013bhd,Yang:2016mvq,Yang:2017shi,Burrello:2020myg}, though there other types of density-dependent
terms usually appearing in the Skyrme forces were considered).
Yet, as in the previous case, they only induce finite corrections to
the $c_2$, $c_4$, $\dots$, parameters, i.e.
\begin{eqnarray}
  \frac{E}{A} + \frac{\delta E}{A} &=& t\,\rho^{2/3} + (c_0 + \delta c_0^R)
  \rho \nonumber \\
  &+&  \left[ c_2 + \delta c_2^R \right] \rho^{5/3}
  + \left[ c_4 + \delta c_4^R \right] \rho^{7/3}
  \nonumber \\
  &+& \dots \, ,
\end{eqnarray}
which again indicates that by choosing the parameters carefully at the first
order at which they appear, their loop corrections can become trivial (modulo
finite terms displaying higher powers of the density not appearing
in the tree level EOS)

\subsection{Power counting}

To determine the power counting one might apply in a first instance
the counting rules of two-body scattering in the vacuum to nuclear matter.
That is, one begins with a strongly interacting two-body system
in which the lowest order contact and the range correction scale as
\begin{eqnarray}
C_0 \sim Q^{-1} \quad \mbox{and} \quad C_2 \sim Q^{-2} \, ,
\end{eqnarray}
which is their usual counting in pionless~\cite{vanKolck:1998bw,Chen:1999tn},
KSW~\cite{Kaplan:1998tg,Kaplan:1998we} and pionful EFT~\cite{Nogga:2005hy,Birse:2005um,Valderrama:2009ei,Valderrama:2011mv,PavonValderrama:2025azr,Long:2011qx,Long:2011xw,Long:2012ve} (though in this latter case, only in the singlet).
Next, the loops are counted as
\begin{eqnarray}
  \int \frac{d^3\vec{q}\,'}{(2\pi)^3}\,
  m_N\,\frac{\theta(k_1'-k_F)\,\theta(k_2'-k_F)}{q^2 - {q'}^2} \sim  Q \, ,
\end{eqnarray}
and then the counting of the higher order contacts can be deduced from
the condition of including the new contacts when they are
needed to absorb the divergences.

This leads to the following naive counting for the different
contributions to the EOS:
\begin{eqnarray}
  \frac{E}{A} &=& \underbrace{t \, \rho^{2/3} + c_0 \, \rho}_{Q^2 ({\rm LO})} + \underbrace{c_2 \, \rho^{5/3}}_{Q^3 ({\rm NLO})} +
  \underbrace{c_4 \, \rho^{7/3}}_{Q^4 ({\rm N^2LO})} \nonumber \\
  &+& \sum_{n = 3}^{\infty} \underbrace{c_{2n} \, \rho^{1 + 2n/3}}_{Q^{n+2} ({\rm N^{n}LO})}
  \, ,
\end{eqnarray}
which is quite simple, though for $C_{2n}$ with $n > 1$
it differs from the vacuum counting.
Indeed, one ends up with $C_4 \sim Q^{-3}$, $C_6 \sim Q^{-4}$, etc.,
instead of the usual $C_{2n} \sim Q^{-2}$ for $n \geq 1$.
Yet, if one were to determine the vacuum power counting from the perturbative
divergence arguments above, one would have also arrived to
the $C_{2n} \sim Q^{-(n+1)}$ scaling.
That is, this might be a limitation of the present argument (as will
be discussed in a few paragraphs), which only considers
the naive divergence structure.
However, it is worth mentioning here that this enhancement of
the $C_{2n}$ couplings coincides with the one proposed by Long and Yang
in pionful EFT on the basis of analyzing their size
in dimensional regularization~\cite{Long:2012ve}.

Irrespective of the scaling of the $C_{2n}$ couplings,
the previous power counting still contains a contradiction:
it implies that many-body perturbation theory must be resummed at
all orders for $C_0$.
Indeed, if a loop counts as $Q$ and $C_0$ counts as $Q^{-1}$,
then the full resummation of $C_0$ becomes necessary.
This is inconsistent with the ${\rm LO}$ form of the EOS proposed here or
with the expected impact of the healing distance in the iteration of
the $C_0$ coupling, as discussed in Section~\ref{sec:healing}.

To understand why this is not the case at finite density,
one begins by considering the explicit expression for the second
order perturbative correction to the EOS:
\begin{eqnarray}
  \frac{E + \delta E}{A} &=&
  t \rho^{2/3} + \frac{1}{2}(1-\frac{1}{g})\rho\,
  \Big[ (C_0 + \delta C_0) \nonumber \\
  &+&
    {\left( \frac{g}{\rho} \right)}^2\,
    I_{0F}(k_F; \Lambda)\,C_0^2 \, + \dots \Big] \, ,
\end{eqnarray}
where the loop integral (averaged for two-body states below the Fermi momentum)
is given by
\begin{eqnarray}
  I_{0F}(k_F; \Lambda) =
  \int_{k_1, k_2 < k_F} \frac{d^3 \vec{k}_1}{(2\pi)^3}
  \frac{d^3 \vec{k}_2}{(2\pi)^3}\,
  \int_{\Lambda} \frac{d^3 \vec{q}\,'}{(2\pi)^3} \,
  \frac{Q_F}{e_{12}} \, ,
  \nonumber \\
\end{eqnarray}
with the ${\Lambda}$ subscript indicating that
the integral has been regularized.
Its explicit evaluation in PDS yields
\begin{eqnarray}
  I_{0F}^{\rm PDS}(k_F; \Lambda) &=& -\frac{1}{(2\pi^2)^2}\Big[
    \frac{m_N}{36 \pi}\, \Lambda \, k_F^6
  \nonumber \\
  && \qquad - \frac{m_N}{210 \pi^2}\,(11 - 2\log{2})\,k_F^7 \Big] \, ,
  \nonumber \\ \label{eq:I0F-PDS}
\end{eqnarray}
and with it the EOS can be rewritten as
\begin{eqnarray}
  \frac{E + \delta E}{A} &=& t\,\rho^{2/3} + \frac{1}{2}(1-\frac{1}{g})\,
  \rho\,\times\, \nonumber \\
  && \Big[
    C_0 + \delta C_0 - C_0^2\,\frac{m_N}{4 \pi}\,(\Lambda - c_{0F} k_F) +
    \dots \Big] \, , \nonumber \\
  \label{eq:delta-EOS}
\end{eqnarray}
with $c_{0F}$ a number, which in PDS is given by
\begin{eqnarray}
  c_{0F} = \frac{6}{35\pi}\,(11-2\log{2}) \approx 0.5246 \, .
\end{eqnarray}
It is worth noting that this number is also the ultraviolet limit of the loop
suppression function $\bar{h}_F(x)$ defined
in Eq.~(\ref{eq:uv-in-medium-loop-modified}).
That is, $\bar{h}_F(0) = c_{0F}$.

If one now considers the RG evolution of the vacuum coupling in pionless EFT
with PDS
\begin{eqnarray}
  C_0(\Lambda) &=& \frac{4\pi}{m_N}\,\frac{1}{\frac{1}{a_0} - \Lambda} \, ,
  \label{eq:C0-running}
\end{eqnarray}
with $a_0$ the scattering length, it is in principle possible to build
an expansion for the EOS around the infrared limit ($\Lambda \to 0$) of
the vacuum ($k_F \to 0$).
For this, one expands the running of the vacuum coupling as
\begin{eqnarray}
  C_0(\Lambda)
  &=& \frac{4 \pi}{m_N} a_0\, +  \frac{4 \pi}{m_N} a_0\,(a_0 \Lambda)
  + \dots
  \nonumber \\
  &=& C_0(\Lambda = 0) + \delta C_0(\Lambda) + \dots \, .
\end{eqnarray}
By explicitly including this expansion in the loop correction
(i.e., in the second line of Eq.~(\ref{eq:delta-EOS})),
one finds that~\footnote{With the intention of keeping the expressions compact
  I will not explicitly write the cutoff at which $C_0(\Lambda)$ is expanded
  into $C_0(\Lambda = \Lambda_0)$ and $\delta C_0$ when evaluating
  the renormalized loop expansion.
}
\begin{eqnarray}
  && C_0 + \delta C_0 - C_0^2\,\frac{m_N}{4 \pi}\,(\Lambda - c_{F0} k_F) + \dots
  \nonumber \\
  && \quad = \frac{4 \pi}{m_N}\,a_0\,
  \left( 1 + c_{0F}\, a_0 k_F + \dots \right) \, ,
  \label{eq:loop-expansion}
\end{eqnarray}
which indicates that the loop is indeed counted as $k_F$ (i.e. $Q$) and thus,
if $a_0 \sim 1/Q$ (strong scattering) the loops must be resummed.
This reproduces the power counting in the vacuum, which
requires the iteration of the contact-range coupling.
Conversely, if $a_0 \sim 1/M$ (weak scattering), it reproduces
the counting for dilute Fermi systems~\cite{Hammer:2000xg}
(as this counting only requires $k_F a_0 \ll 1$ to be valid).

In contrast, if one expands the running of $C_0(\Lambda)$ around
the cutoff $\Lambda^*$:
\begin{eqnarray}
  \frac{1}{C_0(\Lambda)} &=& \frac{m_N}{4\pi}\,(\frac{1}{a_0} - \Lambda)
  \nonumber \\
  &=& \frac{m_N}{4\pi}\,(\frac{1}{a_0} - \Lambda^* + (\Lambda^* - \Lambda))
  \nonumber \\
  &=& \frac{1}{C_0(\Lambda^*)} + \frac{m_N}{4\pi}\,(\Lambda^* - \Lambda)
  \, ,
\end{eqnarray}
then one obtains the renormalized correction
\begin{eqnarray}
  && C_0 + \delta C_0 - C_0^2\,\frac{m_N}{4\pi}\,(\Lambda - c_F\,k_F)
  + \dots  \nonumber \\
  && \quad =
  C_0(\Lambda^*) - C_0^2(\Lambda^*)\,\frac{m_N}{4\pi}\,(\Lambda^* - c_{0F}\,k_F)
  + \dots \, , \nonumber \\
\end{eqnarray}
which implies that the actual size of loop contributions
is $(\Lambda^* - c_{0F}\,k_F)$.
Formally, this still means that the loop scales as $Q$.
Yet, it is apparent that by rewriting 
\begin{eqnarray}
  \Lambda^* = c_{0F}\,k_F^* \, ,
\end{eqnarray}
then the loop corrections vanish for the approximate Fermi momentum
at which the RG flow freezes for the in-medium running of
the coupling:
\begin{eqnarray}
  && C_0 + \delta C_0 - C_0^2\,\frac{m_N}{4\pi}\,(\Lambda - c_F\,k_F)
  + \dots \nonumber \\
  && \quad 
  = C_0(\Lambda^*) - C_0^2(\Lambda^*)\,\frac{m_N}{4\pi}\,c_{0F}\,(k_F^* - k_F)
  + \dots \, . \nonumber \\
\end{eqnarray}
This in turn means that the correct counting of the loops is $(k_F^* - k_F)$,
with $k_F^*$ the Fermi momentum for which the effective contact-range
potential has been optimized.

By choosing the cutoff $\Lambda^*$ what one is doing is accelerating
the EFT expansion of the EOS for $k_F^* = \Lambda^* / c_{0F}$ in PDS,
while for other regulators the only difference is the exact value of
the proportionality constant between $\Lambda^*$ and $k_F^*$.
For the Fermi momentum $k_F^*$ (or its corresponding density $\rho^*$),
the first correction coming from the $C_0$ loops vanishes, while for
Fermi momenta close to this value the loop corrections
will be suppressed by a $(k_F^* - k_F)$ factor.
That is, the EFT expansion is configured in terms of two expansion parameters,
the familiar $Q$ and the new $(k_F^* - k_F)$.

The natural choice for $k_F^*$ in symmetric nuclear matter is around
the saturation point, i.e. $k_F^* \approx 266\,{\rm MeV}$, while
for neutron matter other choices could be tried depending
on which densities one is interested in.
With this, the preferred cutoff in PDS is $\Lambda^* \approx 140\,{\rm MeV}$
at saturation density, where the $C_0$ coupling constant is
\begin{eqnarray}
  C_0(\Lambda^*) &=& \frac{4\pi}{m_N} \frac{1}{\frac{1}{a_0} - \Lambda^*}
  = - \frac{4\pi}{m_N\,\Lambda^*}\,
  \left( 1 + \mathcal{O}(\frac{1}{a_0 \Lambda^*})
  \right) \nonumber \\
  &\approx& -1158\,{\rm MeV}\,{\rm fm}^3 \, ,
\end{eqnarray}
whose value is very similar to the ones usually appearing for the analogous
$t_0$ parameter in the Skyrme potentials.

It might be possible though to derive the strength of the Skyrme parameters
without resort to the equilibrium density, but from the behavior of
the (pionful) wave functions by defining a healing
renormalization scale~\cite{RuizArriola:2016vap}.
This suggests the existence of further RG avenues from which to derive
the saturation point of nuclear matter (instead of using it as
an input) that will be worth exploring in the future.

The extension to the subleading contact-range interactions is not entirely
trivial in regards to power counting.
If one begins with $C_2$, whose running in pionless EFT with PDS is given by
\begin{eqnarray}
  C_2(\Lambda) = \frac{m_N}{4 \pi}\,C_0^2(\Lambda)\,\frac{r_0}{2} \, ,
  \label{eq:C2-running}
\end{eqnarray}
expands around $\Lambda = \Lambda^*$
\begin{eqnarray}
  C_2(\Lambda) &=& \frac{m_N}{4 \pi}\,C_0^2(\Lambda^*)\,\frac{r_0}{2} \
  \nonumber \\
  &+& {\left( \frac{m_N}{4 \pi} \right)}^2\,C_0^3(\Lambda^*)\,r_0\,(\Lambda - \Lambda^*) \nonumber \\
  &+& \mathcal{O}\left( {(\Lambda - \Lambda^*)}^2 \right) \nonumber \\
  &=& C_2(\Lambda^*) + \delta C_2
  + \mathcal{O}\left( {(\Lambda - \Lambda^*)}^2 \right) \, ,
\end{eqnarray}
and then renormalizes the loop correction perturbatively
\begin{eqnarray}
  && C_2 + \delta C_2 - 2\,C_0 C_2\,\frac{m_N}{4\pi}\,(\Lambda - c_{2F}\,k_F)
  + \dots \nonumber \\
  && \quad =
  C_2(\Lambda^*) - 2\,C_0(\Lambda^*)\,C_2(\Lambda^*)\,
  \frac{m_N}{4\pi}\,(\Lambda^* - c_{2F}\,k_F)
  \nonumber \\
  && \quad \phantom{=} + \dots  \, ,
\end{eqnarray}
then it is apparent that the power counting structure of the loops has not
changed, except for a detail: the new numerical constant $c_{2F}$.
Its value depends on the chosen representation for the contacts.
For the momentum-, energy- and density-dependent representations
one obtains, respectively:
\begin{eqnarray}
  c_{2F} &=& \frac{1336 - 192 \log{2}}{378\pi} \approx 1.0130 \, , \\
  c_{2F} &=& \frac{632 - 192 \log{2}}{378\pi} \approx 0.4201 \, , \\
  c_{2F} &=& \frac{6}{35\pi}\,(11-2\log{2}) \approx 0.5246 \, ,
\end{eqnarray}
where for the density-dependent case the factor is identical to the one
when $C_0$ is iterated (i.e. to $c_{0F}$).

This means that in general the Fermi momentum at which the first loop
correction is perfectly suppressed depends
on the order of the coupling (and on the regulator choice and representation
of the contact-range interaction, which is perfectly acceptable
as one is not dealing with an observable).
However, this is not a problem in the sense that the iteration of these
couplings is already suppressed in the first place
owing to their scaling.

Another coupling that deserves a detailed analysis is $C_4$, which
in pionless EFT contains two contributions with different scalings
\begin{eqnarray}
  C_4(\Lambda) &=& \frac{C_2^2(\Lambda)}{C_0(\Lambda)} +
  \frac{m_N}{4\pi}\,C_0^2(\Lambda)\,v_2  \nonumber \\
  &=& C_4^A(\Lambda) + C_4^B(\Lambda) \, ,
\end{eqnarray}
where $C_4^A$ absorbs the cutoff dependence from the iteration of $C_2$,
while $C_4^B$ contains the physical information
about the shape parameter $v_2$.
As is well known, in the vacuum $C_4^A$ enters at ${\rm N^2LO}$ and
$C_4^B$ at ${\rm N^3LO}$.
Owing to the linear nature of the contributions, it is possible to analyze
their loop corrections separately.
For $C_4^A$, by repeating the same steps as for $C_2$, one has
\begin{eqnarray}
  && C_4^A + \delta C_4^A - [C_2^2 + 2\,C_0 C_4^A]\,\frac{m_N}{4\pi}\,(\Lambda - c_{4F}^A\,k_F) + \dots \nonumber \\
  && \quad =
  C_4^A(\Lambda^*) - \Big[  C_2^2(\Lambda^*) \nonumber \\
  && \quad \quad \,\, + 2\,C_0(\Lambda^*)\,C_4^A(\Lambda^*) \Big]\,
  \frac{m_N}{4\pi}\,(\Lambda^* - c_{4F}^A\,k_F) \nonumber \\
  && \quad + \dots  \, ,
\end{eqnarray}
while for $C_4^B$
\begin{eqnarray}
  && C_4^B + \delta C_4^B - 2\,C_0 C_4^B\,\frac{m_N}{4\pi}\,(\Lambda - c_{4F}^B\,k_F) + \dots \nonumber \\
  && \quad =
  C_4^B(\Lambda^*) - 2\,C_0(\Lambda^*)\,C_4^B(\Lambda^*)\,
  \frac{m_N}{4\pi}\,(\Lambda^* - c_{4F}^B\,k_F) \nonumber \\
  && \quad + \dots  \, ,
\end{eqnarray}
with $c_{4F}^A$ and $c_{4F}^B$ new numerical constants.
In the momentum-dependent representation of the couplings their values are
\begin{eqnarray}
  c_{4F}^A &=& \frac{5636 - 564\,\log{2}}{891 \pi} \approx 1.874 \, , \\
  c_{4F}^B &=& \frac{6329 - 564\,\log{2}}{891 \pi}
  \approx 2.121 \, ,
\end{eqnarray}
in the energy-dependent representation one has
\begin{eqnarray}
  c_{4F}^A &=& c_{4F}^B = \frac{1337 - 564\,\log{2}}{891 \pi} \approx 0.3380 \, ,
\end{eqnarray}
and finally, in the density-dependent representation they take their most
simple form:
\begin{eqnarray}
  c_{4F}^A &=& c_{4F}^B = c_{0F} \, ,
\end{eqnarray}
With the exception of the density dependent couplings, it is now evident that
the cutoff asynchrony of the iteration of higher derivative couplings
increases with the order.

Independently of the representation, the present analysis indicates that,
in principle, the two contributions to $C_4$ are counted differently
not only in the vacuum , but in nuclear matter as well.
That is, the scalings still are
\begin{eqnarray}
C_4^A \sim \frac{1}{Q^3} \quad \mbox{and} \quad C_4^B \sim \frac{1}{Q^2} \, .
\end{eqnarray}
It is necessary to emphasize {\it in principle} because
this counting is only possible provided one includes
counterterms of the $C_4^A$ type
in the first place.
That is, counterterms that do not carry physical information
but are there to remove the residual cutoff dependence.
This type of counterterm has been explicitly identified
in~\cite{Valderrama:2016koj,Valderrama:2019yiv,PavonValderrama:2026pmj},
where they have been called redundant or auxiliary.

To summarize, the features of the power counting of
two-body contact-range forces in nuclear matter are:
\begin{itemize}
\item[(i)] The couplings scale {\it a priori} in exactly the same way
  as in the vacuum: the ${\rm LO}$ coupling is counted as $1/Q$.
  However, for subleading couplings this requires the inclusion of auxiliary
  couplings removing the residual cutoff dependence. Without them
  certain subleading couplings will require promotions.
\item[(ii)] If the couplings are evaluated at the cutoff $\Lambda = \Lambda^*$,
  then their iteration in a loop counts as a factor of $(k_F^* - k_F)$,
  with $k_F^*$ a Fermi momentum proportional to $\Lambda^*$
  (where the proportionality constant also depends on the number of
  derivatives of the contact-range interaction) 
\item[(iii)] Formally this is compatible with how loops are counted
  in the vacuum (i.e., as $Q$), except for the fact that in the medium
  loops will be suppressed at Fermi momenta close to $k_F^*$.
\item[(iv)] The practical consequence if that there is a Fermi momentum
  $k_F^*$ in the vicinity of which the iterations of the ${\rm LO}$
  coupling (evaluated at the cutoff $\Lambda^*$) are suppressed.
\item[(v)] For subleading couplings this is also the case, though
  their $k_F^*$ might be different than for the ${\rm LO}$ coupling.
  Nonetheless, this is inconsequential as their iterations are
  suppressed regardless of whether loops are counted
  as $Q$ or $(k_F^* - k_F)$.
\end{itemize}
It is important to notice here that $\Lambda^*$ is not really a cutoff,
but the healing scale for nuclear matter with Fermi momentum $k_F^*$.
By tuning $\Lambda^*$ one is optimizing the EFT
as to describe nuclear matter at a given density.

Yet, the $(k_F^* - k_F)$ counting rule is by itself the consequence of
structuring nuclear matter calculations around the vacuum
contact-range couplings (whose running does not depend on $k_F$)
instead of the in-medium couplings (whose running
depends on $k_F$), as already discussed at
the end of Section~\ref{sec:healing}.

\subsection{Derivative and primitive countings}

There are two possibilities regarding the power counting of subleading order
terms, both of which renormalize the perturbative
corrections to the energy per nucleon.
One is a primitive or bottom-up approach, where new couplings are included
wherever a divergence enters in this perturbative expansion.
In pionless EFT  this leads to the scaling $C_{2n} \sim 1/Q^{n+1}$
in the medium.

Another is a derivative or top-down approach, where the starting point is
the power counting of the couplings in the vacuum, and which explicitly
requires the inclusion of auxiliary or redundant counterterms.
In the vacuum this type of counterterm absorbs residual cutoff
dependence, that is, cutoff dependence that vanishes
in the hard cutoff limit.
In the medium, if one expands around the MF limit (which corresponds
to the infrared limit of the renormalization group), this residual
cutoff dependence is no longer residual and manifests
as divergences~\footnote{This might seem confusing at first sight, yet
  the underlying idea is relatively simple: first, the infrared limit should be
  thought as the rules by which to derive the EFT expansion one is interested
  in (e.g. the MF limit). Once the rules are known, it is possible to check
  for cutoff independence within the expansion already derived
  from the infrared limit.}.
Thus, the inclusion of auxiliary counterterms turns out to be crucial
if the original power counting of the two-nucleon system
in the vacuum is to be kept.

This point might be illustrated with the specific example of the first
iteration of the range correction to the T-matrix in the vacuum:
\begin{eqnarray}
  \delta T = \frac{C_2^2 I_0 k^4}{{(1 - C_0 I_0)}^3} +
  \frac{C_4^A k^4}{{(1 - C_0 I_0)}^2} \, ,
\end{eqnarray}
with $C_0$, $C_2$ couplings and $I_0 = I_0(k; \Lambda)$ the loop integral.
Note the inclusion of an auxiliary term proportional to the coupling $C_4^A$
that is there to absorb the residual cutoff dependence.
The explicit evaluation of $\delta T$ in PDS yields  
\begin{eqnarray}
  \delta T
  &=& \frac{4\pi}{m_N}\frac{1}{{(\frac{1}{a_0} + i k)}^3}\,
  \Bigg[ {\left( \frac{1}{2} r_0 k^2 \right)}^2\,\frac{\Lambda + i k}{\Lambda - \frac{1}{a_0}}
        \nonumber \\
    &&
    + \frac{m_N}{4 \pi}\,(\frac{1}{a_0} + i k)\,
    {\big( \Lambda - \frac{1}{a_0} \big)}^2\,C_4^A\,k^4
    \Bigg] \, ,
\end{eqnarray}
where the second line contains only the auxiliary term.
If one interprets the PDS renormalization scale $\Lambda$ as a cutoff, or
alternatively, if one treats the calculation in PDS as a simplified
calculation in cutoff regularization, it is apparent that
as $\Lambda \to \infty$ the contribution from the iteration of the range
correction is finite and converges to the value it should be expected to
have from the expansion of range corrections in the T-matrix:
\begin{eqnarray}
  T
  &=& - \frac{4\pi}{m_N}\,\frac{1}{-\frac{1}{a_0} + \frac{1}{2}r_0 k^2 - i k}
  \nonumber \\
  &=& + \frac{4 \pi}{m_N}\,\frac{1}{\frac{1}{a_0} + i k}\,
  \sum_{n=0}^{\infty} {\left(
    -\frac{\frac{1}{2}r_0 k^2}{\frac{1}{a_0} + i k} \right)}^n \, .
  \nonumber \\
\end{eqnarray}
That is, $C_4^A$ is unnecessary for the $\Lambda \to \infty$ limit
to be well-defined.
Of course, if included, the residual cutoff dependence disappears in PDS.

However, if one were to re-expand all the loops,
one would find instead
\begin{eqnarray}
  \delta T_{\rm exp}
  &=& \left[ C_4^A\, + C_2^2 I_0\,\sum_{n=0}^{\infty} {(C_0 I_0)}^n
    \right] k^4 \,{\left( \sum_{n=0}^{\infty} {(C_0 I_0)}^n \right)}^2
  \nonumber \\
  &=& \left[ C_4^A - C_2^2 \frac{m_N}{4 \pi} (\Lambda + i k)
    \right] k^4 \, \Big(  1 + \mathcal{O}{(C_0 I_0)} \Big) \, ,
  \nonumber \\
\end{eqnarray}
where by considering the expansion of $C_2(\Lambda)$ around $\Lambda = 0$
it becomes evident that $C_4^A$ now cures a divergence,
a different role from the one it originally had
in the calculation where the $C_0$ bubbles
were resummed.

While this is but a curiosity in the two-body sector, in nuclear matter it
becomes crucial, at least if one wants a microscopic derivation of the EOS
and provided one is expanding around the MF approximation, which forces
the re-expansion of the $C_0$ bubbles as in the example above.
The issue is that most two-body EFT descriptions ignore the existence of
auxiliary counterterms, as they are not required to ensure
renormalizability, understood usually as the existence of
the ultraviolet limit.
The only exception are improved
actions~\cite{Contessi:2023yoz,Contessi:2024vae,Contessi:2025xue},
though they are currently limited to ${\rm LO}$.

\subsection{Pion exchanges}

While there are clean arguments of why the loops of the contacts
are suppressed at finite densities (or, at least, at a given
density), it is not evident if this is also the case 
for pion exchanges.
There are reasons for their suppression too, though they belong to
the category of educated guesses rather than that of concrete proof.

The question can be narrowed down to what happens with the pion loops
in pionful EFT.
In pionless there are no pion exchanges, while in KSW pion exchanges
are counted as in naive dimensional analysis, that is:
\begin{eqnarray}
  V_{F} = \underbrace{V_{\rm OPE}}_{Q^0} + \underbrace{V_{\rm TPE(L)}}_{Q^2} +
  \underbrace{V_{\rm TPE(SL)}}_{Q^3} + \dots \, ,
\end{eqnarray}
where OPE refers to one-pion exchange, and TPE(L) and (SL) to leading
and subleading two-pion exchange.
Thus, pion contributions enter at ${\rm NLO}$ in the EOS.

If one now considers pionful EFT (or, more properly, the pionful EFT
where OPE is non-perturbative), consistency requires that pion
corrections to the running of the contact-range couplings
follow their iteration pattern.
That is, a coupling in pionful EFT receives a contribution from OPE at tree
level, then a one-loop OPE contribution and a one-loop contribution
from itself, and so on {\it ad infinitum}.
Thus, if one wants the expansion of the couplings in nuclear
matter to be consistent with the pion exchanges,
the latter should in principle follow
the iteration pattern of the former.

This issue is intertwined with the fact that what matters at finite
density is the infrared running of the couplings for $\Lambda \to \Lambda^*$.
With cutoffs $\Lambda^*$ of the order of the pion mass at equilibrium density,
it happens that pions are effectively perturbative at
these regularization scales.

Besides, it is worth noticing that the running of the couplings
in the presence of the OPE potential can be reproduced perturbatively
until the appearance of the first deeply bound state generated by OPE.
The reason is that perturbation theory for an attractive singular,
power-law potential converges to the two-body wave function
beyond its first zero (see~\cite{PavonValderrama:2026pmj}
for a concrete example using the van der Waals potential).
Owing to the fact that the couplings can be calculated from the form of the
wave function (for instance, using Eq.~(\ref{eq:C0-running-delta-shell})
in the particular case of a delta-shell regulator),
this means that the running of the coupling can be reproduced
until the first zero of the wave function, at which point
the non-perturbative coupling hits a singularity
thus signaling the failure of
perturbation theory.

Yet, for connecting the few-body description of nuclear forces with
the properties of nuclear matter, it is enough to reproduce
the running of the couplings in the vicinity of $\Lambda^*$,
which does not require higher order perturbation theory.
If high enough orders are involved though, matching the vacuum and
the in-medium renormalizations will require the addition of
auxiliary counterterms simply because the divergence
structures are not identical for the perturbative and
non-perturbative calculations.

But while perturbation theory for the OPE potential converges
for the couplings, this does not imply its convergence
for two-nucleon scattering observables.
There perturbative pions fail at center-of-mass momenta of the order of
$(100-200)\,{\rm MeV}$, at which the tensor part of the OPE potential
requires a non-perturbative treatment~\cite{Fleming:1999ee,Birse:2005um}.
It is not clear though how this limitation manifests at finite density.
For the EOS of nuclear matter the loop integrals are evaluated at
external momenta below $k_F$ while the internal momenta
over which the loop runs are above $k_F$.
This by itself does not imply that tensor OPE becomes perturbative
in nuclear matter, but is a strong hint in that direction.

\subsection{Breakdown scale}

There exists a clear analogy between scattering in the two-nucleon system
and the perturbative expansion of the energy per nucleon
in terms of the two-nucleon potential.
They respectively involve the expansions of the T- and G-matrices,
for which the only substantive difference is the inclusion of
the projector $Q_F$ in the latter.
In particular, they are completely analogous in what regards
the distinction of long- and short-range physics.
If one were to consider a generic two-body potential containing a long- and
short-range component
\begin{eqnarray}
  V = V_L + V_S \, ,
\end{eqnarray}
the two-potential formulas for the T- and G-matrices are formally equivalent
\begin{eqnarray}
  T &=& T_S + ( 1 + T_S \,\frac{1}{E - H_0} ) \hat{T}_L (\frac{1}{E-H_0}\, T_S + 1 ) \, , \nonumber \\ \\
  G &=& G_S + ( 1 + G_S\, \frac{Q_F}{E - H_0}) \hat{G}_L
  (\frac{Q_F}{E - H_0}\,G_S + 1 ) \, , \nonumber \\
\end{eqnarray}
with $T_S$ and $\hat{T}_L$ defined as
\begin{eqnarray}
  T_S &=& V_S + V_S\,\frac{1}{E-H_0}\,T_S \, , \\
  \hat{T}_L &=& V_L + V_L\,\frac{1}{E - H_0 - V_S}\,\hat{T}_L \, , 
\end{eqnarray}
and $G_S$ and $\hat{G}_L$ defined analogously,
but with a $Q_F$ projector over every
energy denominator.
The point is that the expansions of $T$ and $G$ in powers of $m_L/m_S$,
or more generally light over heavy scales, will be really similar
in both cases, with the main difference being that $G$ contains
$k_F$ and the total momentum of the two fermions
as new light scales.
Notice that this idea has been previously discussed
in~\cite{Furnstahl:1999ix} by means of a concrete
calculation in a contact-range theory.

However, though the EOS depends on the G-matrix and its expansion, this
does not imply that the analogy between the T- and G-matrix translates
into similar expansion parameters for two-body scattering
and the EOS.
Here one has to explicitly check that the averaging over states
within the Fermi sea does not alter the properties of
the expansion.

A more transparent analogy between two-body scattering and the two-body
potential energy per nucleon is provided by considering the perturbative
effects of the short-range potential
\begin{eqnarray}
  V_S(\vec{q}\,) \propto \frac{g_S^2}{m_S^2 + {\vec{q}\,}^2} \, ,
\end{eqnarray}
which corresponds to the exchange of a heavy meson of mass $m_S$.
The perturbative effect on $\cot{\delta(k)}$ (with $\delta(k)$ the phase shift)
is
\begin{eqnarray}
  \delta (\cot{\delta}) = m_N k\,\langle \Psi_{k} | V_S | \Psi_{k} \rangle \, ,
\end{eqnarray}
with $| \Psi_{k} \rangle$ the two-body wave function in a normalization
that asymptotically approaches the S-wave projection of
$| \vec{k}\, \rangle$.
Irrespective of the details of the wave function, one will find a contribution
behaving as
\begin{eqnarray}
  \delta (\cot{\delta(k)}) \propto \log{(1 + \frac{4 k^2}{m_S^2})} \, ,
\end{eqnarray}
which implies the existence of a left branch cut starting a
$k = \pm i (m_S/2)$ and hence that the Taylor expansion of $\cot{\delta}$
in powers of $k$ only converges for $k < m_S/2$.
From a direct comparison with the EFT expansion for $\cot{\delta}$
\begin{eqnarray}
  \cot{\delta(k)} = \sum_{\nu = 0}^{\infty} [ \cot{\delta}(k) ]^{(\nu)}\,
  {\left( \frac{Q}{M} \right)}^{\nu} \, ,
\end{eqnarray}
and the observation that the momentum $k$ is a light scale, 
one arrives at the conclusion that $M = m_S / 2$.

If one now considers the energy per nucleon, the perturbative contribution
from $V_S$ is given by
\begin{eqnarray}
  \frac{\delta E}{A} = \frac{\Omega}{2 A}\,
  \int_{k_1, k_2 < k_F} \frac{d^3 \vec{k}_1}{(2\pi)^3} \frac{d^3 \vec{k}_2}{(2\pi)^3}\,
  \langle V_{S} \rangle \, ,
\end{eqnarray}
with $\langle V_S \rangle$ denoting the projection of $V_S$
into an antisymmetric two-body state and
its sum on spin / isospin.
The contribution to the energy per particle of this correction contains a term
proportional to
\begin{eqnarray}
  \frac{\delta E}{A} \propto \log{(1 + \frac{4 k_F^2}{m_S})} \, ,
\end{eqnarray}
which has a branch cut at $k_F = \pm i (m_S/2)$.
Thus, from this argument the EFT expansions for two-body scattering in powers
of the external momentum $k$ and the EOS in powers of $k_F$ work in exactly
the same way: they are expansions in $k_F/M$ and $k/M$, where the breakdown
scale $M$ is identical, as originally argued in~\cite{Furnstahl:1999ix}.

The literal identification of $m_S$ with the rho meson mass
($m_{\rho} \approx 770\,{\rm MeV}$) implies
$M = m_{\rho}/2 \approx 385\,{\rm MeV}$ for two-body scattering
and nuclear matter.
This in turn leads to a breakdown density of $\rho_M \approx 3\,{\rho}_0$,
with $\rho_0$ the equilibrium density of symmetric nuclear matter.

Though this breakdown scale is plausible, it might err
on the conservative side~\footnote{Or it might err on the optimistic
  side: the EFT expansion is most probably an asymptotic expansion with
  a zero radius of convergence when all terms are included (yet seemingly
  convergent for the first few terms). If this is the case the breakdown scale
  should be understood with respect to the apparent convergence properties
  of the lower order terms of the EFT expansion.
}.
This argument, when applied to pionless EFT, results in a breakdown
scale of $M = m_{\pi}/2 \approx 70\,{\rm MeV}$.
Common pionless EFT practice suggests $M = m_{\pi}$ instead, while
the competing Bayesian analysis in~\cite{Ekstrom:2024dqr}
results in $M \approx 200\,{\rm MeV}$. 
And though there is a Bayesian analysis~\cite{Bub:2024gyz} that generates $M$
close to $m_{\pi} / 2$, it uses NDA instead of the power counting of
pionless EFT.
This indicates that $M = m_{\pi}/2$ (and hence $M = m_{\rho}/2$) might
result in an underestimation of the actual breakdown scale.

Bayesian estimations in EFT-inspired potentials containing pions (but
following NDA instead of the pionful EFT counting) yield
$M \approx (400-600)\,{\rm MeV}$ depending on the specific
two-nucleon potential studied~\cite{Furnstahl:2015rha,Melendez:2017phj}.
This range is rather wide and when translated into nuclear matter leads
to a breakdown density of $\rho_M \approx (3-12)\,{\rho}_0$.
It is interesting to mention that the recent methodology of~\cite{Millican:2024yuz,Millican:2025sdp} points to $M \approx (600-750)\,{\rm MeV}$,
potentially leading to $\rho_M \approx (12-24)\,{\rho}_0$.

There is a different argument which comes from the observation that
$k_F^*$ is defined from what might be called
the healing cutoff $\Lambda^*$.
This implies that the hardest $k_F^*$ around which the EOS can be expanded
is determined by the hardest $\Lambda^*$ that can be realistically chosen,
which in turn is given by the appearance of the first pole (i.e. the first
deeply bound state) in the running of the full,
non-perturbative calculation of $C_0(\Lambda)$.
That is, if $C_0(\Lambda)$ diverges at $\Lambda = \Lambda_B$
then $\Lambda^* < \Lambda_B$ is required, from which
the breakdown scale for $k_F^*$ can be derived once
the corresponding proportionality factor is included.

The exact location of $\Lambda_B$ is not known because it depends
on the full EFT potential (if such a thing exists) for which
the expansion becomes more and more computationally
expensive with each new order.
If one includes the finite-range physics up to subleading TPE, then
$\Lambda_B$ typically sits around $500\,{\rm MeV}$, give or take.
For a sharp-cutoff the proportionality is given by
$k_F^* = ({2}/{\pi})\,({\Lambda^*}/{c_{0F}})$ ($\approx 1.2136 \Lambda^*$),
which leads to $k_B \approx 600\,{\rm MeV} $
for $\Lambda_B \approx 500\,{\rm MeV}$, thus implying a
breakdown density of $\rho_M \approx (11-12)\,\rho_0$.

\subsection{Three- and four-body forces}

In symmetric nuclear matter, owing to the existence of four internal
degrees of freedom, besides the already discussed 2-body forces
there are also 3- and 4-body forces.
For simplicity I will only consider density-dependent few-body forces
\begin{eqnarray}
  V_{3B} &=& \sum_{n \geq 0} D_{2n} k_F^{2n} \, , \\
  V_{4B} &=& \sum_{n \geq 0} E_{2n} k_F^{2n} \, , 
\end{eqnarray}
where $D_{2n}$, $E_{2n}$ are the coupling constants.
With these additions the EOS at tree level reads
\begin{eqnarray}
  \frac{E}{A} &=& t \rho^{2/3} + c_0 \rho + c_2 \rho^{5/3} +
  \sum_{n \geq 2} c_{2n} \rho^{1+2n/3} \nonumber \\
  &+& d_0 \rho^2 + d_2 \rho^{8/3} + \sum_{n \geq 2} d_{2n} \rho^{2+2n/3} 
  \nonumber \\
  &+& e_0 \rho^3 + e_2 \rho^{11/3} + \sum_{n \geq 2} e_{2n} \rho^{3+2n/3}  \, ,
\end{eqnarray}
where the $d_{2n}$ and $e_{2n}$ coefficients are proportional to
the $D_{2n}$ and $E_{2n}$ couplings.

If one momentarily ignores the interference with the two-body forces
(which will be important), the loop corrections to the 3-body
contributions is simply given by the first iteration of
the 3-body force
\begin{eqnarray}
  \frac{\delta E_3}{A} &\propto& \frac{1}{\rho}\,\int_{k_1, k_2, k_3 < k_F}\,
  \frac{d^3 \vec{k}_1}{{(2\pi)}^3} \,
  \frac{d^3 \vec{k}_2}{{(2\pi)}^3} \,
  \frac{d^3 \vec{k}_3}{{(2\pi)}^3} \nonumber \\
  && \times
  \int
  \frac{d^3 \vec{p}_{12}\,'}{{(2\pi)}^3}
  \frac{d^2 \vec{p}_3\,'}{{(2\pi)}^3}\,
  \frac{|V_{3B}|^2}{e_{123}}\,Q_F\, ,
\end{eqnarray}
with the energy denominator 
\begin{eqnarray}
  e_{123}
  &=& \sum_i \frac{\vec{k}_i^2}{2m_N} - \sum_i \frac{{\vec{k}_i'\,}^2}{2m_N} 
  \\
  &=& \frac{{{p}_{12}}^2}{m_N} + \frac{3\,{p_3}^2}{4\,m_N} -
  \frac{{{p}_{12}'}^2}{m_N} - \frac{3\,{p_3'}^2}{4\,m_N} \, ,
\end{eqnarray}
where the second line is written in terms of the Jacobi momenta.
The projector $Q_F$ is now defined as
\begin{eqnarray}
  Q_F &=& \theta (k_1' - k_F)\,\theta (k_2' - k_F)\,\theta (k_3' - k_F) \, ,
\end{eqnarray}
with
\begin{eqnarray}
  \vec{k}_1' &=& \frac{\vec{P}\,'}{3} - \frac{\vec{p}_{3}\,'}{2}
  + \vec{p}_{23}\,'
  \, , \\
  \vec{k}_2' &=& \frac{\vec{P}\,'}{3} - \frac{\vec{p}_{3}\,'}{2}
  - \vec{p}_{23}\,'
  \, , \\
  \vec{k}_3' &=& \frac{\vec{P}\,'}{3} + \vec{p}_{3}\,' \, .
\end{eqnarray}
Owing to momentum conservation, $\vec{P}\,' = \vec{P}$,
with $\vec{P} = \vec{k}_1 + \vec{k}_2 + \vec{k}_3$
the total momentum of the three initial
fermions in the sea.

The degree of divergence of this integral can be determined from defining
a six dimensional momentum variable
${p'}^2 = {p_{12}'}^2 + \frac{3}{4}\,{{p}_3'}^2$,
yielding
\begin{eqnarray}
  && \int_{\Lambda}
  \frac{d^3 \vec{p}_{12}\,'}{{(2\pi)}^3}
    \frac{d^3 \vec{p}_3\,'}{{(2\pi)}^3}\,
  \frac{|V_{3B}|^2}{e_{123}}\,Q_F\, \nonumber \\
  && \quad \sim m_N\,|V_{3B}|^2\,
  \int_{\Lambda} \frac{d^6 \vec{p}\,'}{(2\pi)^6} \frac{1}{{p'}^2} \nonumber \\
  && \quad \sim m_N\,|V_{3B}|^2\,\Lambda^4 \, ,
\end{eqnarray}
where $\Lambda$ is the cutoff (that has been directly
applied to the six dimensional momentum).
Naively, this implies that the iteration of the 3-body force diverges as
\begin{eqnarray}
  \frac{\delta E_3}{A}
  &\propto& m_N \Lambda^4\,
  \left[ D_0^2 + 2 \, D_0 D_2\, k_F^2 + \dots \right] \, , 
\end{eqnarray}
while for the 4-body force (by repeating similar arguments
and assumptions) the divergence is
\begin{eqnarray}
  \frac{\delta E_4}{A} &\propto&
  m_N \Lambda^7\,\left[ E_0^2 + 2 \,E_0 E_2\, k_F^2 + \dots \right] \, .
\end{eqnarray}

The naive reading of this divergence structure suggests that the subset of
couplings that is closed under loop corrections is composed
by the lowest order 3- and 4-body forces.
Yet, at this point this is but a conjecture and far from being proven:
it only takes into account the leading divergence $\Lambda^{3 A - 5}$
in the loop of the $A$-body force, which is followed by density-dependent
divergences of the type $\Lambda^{3 A - 5 - 2 n}\,k_F^{2n}$ that are
much more difficult to handle if one wants to show closure.
Here it becomes necessary to take into account the interference with
the two-body forces, particularly in the 3-body case
(as happens, for instance, in dilute systems~\cite{Hammer:2000xg},
though there one goes from a divergence involving
two-body forces to the inclusion of 3-body forces).
After all, the 3-body contact appears from the necessity of preventing
the Thomas collapse induced by the contact-range two-body forces.

For the specific case of the 3-body forces, the complete divergent
structure of the 3-body loop actually is
\begin{eqnarray}
  \frac{\delta E_3}{A}
  &\propto& m_N \left( c_4\,\Lambda^4 + c_2\,k_F^2 \Lambda^2 + c_0\,k_F^4\,
  \log{\frac{\Lambda}{k_F}} \right) \nonumber \\
  &\times& \,\left[ D_0^2 + 2 \,D_0 D_2 k_F^2 + \dots \right] \, , 
\end{eqnarray}
which is not renormalizable with the $\delta D_0$ correction to
the 3-body coupling alone (where it is interesting to notice that
the vacuum version of this problem has been already discussed
in~\cite{Epelbaum:2018zli}).
Here the inclusion of auxiliary $\delta D_2$ and $\delta D_4$ corrections
(without the corresponding finite part) should in principle be able to absorb
the $k_F$-dependent divergences.

Yet, a more satisfying solution might be found in the observation
that the iteration of the two-body coupling $C_0$ is only
suppressed when sandwiched and averaged between states
in the Fermi sea.
Within the $D_0^2$ loop it is thus perfectly possible to nest $C_0$
two-body loops, which (if not suppressed) end up reproducing
the in-medium 3-body wave function~\footnote{To understand this step one begins
  by noticing that the sum of the two-body bubbles
  reproduces the 3-body G-matrix (or the part of it
  that does not depend on the 3-body forces):
  \begin{eqnarray}
    && V_{3B} \frac{Q_F}{E-H_0} \left[ 1 + \sum_{i<j}\,
      {\left( V_{2B} \right)}_{ij} \frac{Q_F}{E-H_0} +
      \dots \right] V_{3B} \nonumber \\
    && \qquad = V_{3B} \frac{Q_F}{E-H_0}\,
    \left[ 1 +  \left( \sum_{k}\,G^{(k)} \right) \frac{Q_F}{E-H_0}
      \right]\,V_{3B} \, , \nonumber
  \end{eqnarray}
  with $G = \sum_k G^{(k)}$ the G-matrix expressed as the sum of its Faddeev
  components.
  From the relation
  $[ 1 + ({{Q_F}}/{(E-H_0)}) G(E) ]\,| \phi \rangle =
  | \Psi \rangle $
  with $| \phi \rangle$ and $| \Psi \rangle$ the free and full wave
  functions, respectively, one arrives at Eq.~(\ref{eq:3B-loop-full}).
  Strictly speaking, this is not the full 3-body wave function as it does
  not contain 3-body contributions.
  But for the contact-range case the difference is an ambiguity in the
  value of the scale $\Lambda_3$ appearing in Eq.~(\ref{eq:3b-wavefunction-a}),
  which does not affect the divergence structure of the theory.
}:
\begin{eqnarray}
    && \int_{\Lambda}
  \frac{d^3 \vec{p}_{12}\,'}{{(2\pi)}^3}
    \frac{d^3 \vec{p}_3\,'}{{(2\pi)}^3}\,
    \frac{|V_{3B}|^2}{e_{123}}\,Q_F \nonumber \\
    && \quad \to
    \int_{\Lambda}
    \frac{d^3 \vec{p}_{12}\,'}{{(2\pi)}^3}
    \frac{d^3 \vec{p}_3\,'}{{(2\pi)}^3}\,
    \frac{|V_{3B}|^2}{e_{123}}\,Q_F\,\Psi_{3F}(\vec{p}_{12}\,', \vec{p}_3\,',\vec{P}) \, .
    \nonumber \\
    \label{eq:3B-loop-full}
\end{eqnarray}
Assuming that at short distances (or high momenta) the in-medium wave
function behaves as the vacuum one, then
\begin{eqnarray}
  \Psi_{3F}(\vec{p}_{12}, \vec{p}_3 , \vec{P})
  \to \frac{a(p_3)}{p_{12}^2 + \frac{3}{4}\,p_{3}^2} \, ,
\end{eqnarray}
with
\begin{eqnarray}
  a(p_3) \to
  \frac{1}{p_3^2}\,\sin{\left[ s_0\,\log{\frac{p_3}{\Lambda_3}} \right]} \, ,
  \label{eq:3b-wavefunction-a}
\end{eqnarray}
where $s_0 \approx 1.00624 $ for a system of three identical particles
in S-wave, and with $\Lambda_3$ a momentum scale that is fixed
by the 3-body force (check~\cite{Bedaque:1998kg,Bedaque:1998km,Bedaque:1999ve}
for further details).
From this, the divergence structure might be schematically written as
\begin{eqnarray}
    && 
    \int_{\Lambda}
    \frac{d^3 \vec{p}_{12}\,'}{{(2\pi)}^3}
    \frac{d^3 \vec{p}_3\,'}{{(2\pi)}^3}\,
    \frac{|V_{3B}|^2}{e_{123}}\,Q_F\,
    \Psi_{3F}(\vec{p}_{12}\,', \vec{p}_3\,' , \vec{P}\,) 
    \nonumber \\
    && \qquad \sim m_N\,D_0^2\,\Big(
    c_0\,\,f_0(s_0\log{\frac{\Lambda}{\Lambda_3}}) + \dots 
    \Big) \, , 
\end{eqnarray}
with $f_0(x)$ an oscillatory function (whose value is bounded, and thus
this is not really a divergence but an ambiguity) and
where the dots indicate finite (or non-ambiguous) contributions.
Now the $\delta D_0$ correction is able to absorb the cutoff ambiguity,
from which closure follows for the lowest order 3-body force.
Meanwhile, repeating the previous arguments for the 4-body force leads to
a linear divergence (times an oscillatory function) that can be absorbed
by the $\delta E_0$ correction.

Thus, from the closure argument, one obtains the EOS:
\begin{eqnarray}
  \frac{E}{A} \Big|_{\rm closed} = t \rho^{2/3} + c_0 \rho + d_0 \rho^2 + e_0 \rho^3 \, ,
\end{eqnarray}
where it is important to stress that this is not necessarily the ${\rm LO}$
EOS, but a possible candidate to the ${\rm LO}$ EOS.
It is not the most general candidate either:
as will be explained in Section \ref{sec:density-dependence}, there are
density-dependent terms that originate from integrating out intermediate
range physics that have to be included in the EOS.
These terms also make unnecessary the 3-body force in its role of
avoiding the collapse of nuclear matter (though they do not prevent
the inclusion of 3-body forces if they are still required
by power counting arguments).

For illustrating the power counting of the 3- and 4-body contributions
with a specific example, I will consider their scaling
in pionless EFT~\cite{Bedaque:1998kg,Bedaque:1998km,Bedaque:1999ve}
\begin{eqnarray}
  D_0 \sim Q^{-4} \,\, \mbox{(${\rm LO}$)} \quad \mbox{and} \quad
  E_0 \sim Q^{-6} \,\, \mbox{($\rm NLO$)} \, ,
\end{eqnarray}
where for the 4-body force I have followed its conjectured counting
in~\cite{Bazak:2018qnu}.
The subleading 3- and 4-body forces inherit the scaling of their ${\rm LO}$
counterparts
\begin{eqnarray}
  D_{2n} \sim Q^{-4} \quad \mbox{and} \quad
  E_{2n} \sim Q^{-6} \, ,
\end{eqnarray}
where the first of these scalings is a consequence of sandwiching the 3-body
contact with the 3-body wave function (as done
in~\cite{Valderrama:2016koj}), while the second is the naive
expectation once one considers that $E_0$ is already a perturbative
correction (as normally all derivative couplings are equally promoted).
Meanwhile, the pure 3- and 4-body loops scale as $Q^4$ and $Q^7$ (that is,
the same power they diverge with).
With this as the starting point, one ends up with
\begin{eqnarray}
  \frac{E_3}{A} &=& \underbrace{d_0 \, \rho^2}_{Q^2 ({\rm LO})} +
  \underbrace{d_2 \,\rho^{8/3}}_{Q^4 ({\rm N^2LO})} + \underbrace{d_4 \,\rho^{10/3}}_{Q^6 ({\rm N^4LO})} + \dots \, , \\
  \frac{E_4}{A} &=& \underbrace{e_0 \, \rho^3}_{Q^3 ({\rm NLO})} +
  \underbrace{e_2 \, \rho^{11/3}}_{Q^5 ({\rm N^3LO})} + \underbrace{e_4 \, \rho^{14/3}}_{Q^7 ({\rm N^5LO})} + \dots \, ,
\end{eqnarray}
for the derivative (or top-down) counting.
For other countings the arguments are analogous, though in pionful EFT
there is the open problem of how to count the 3- and 4-body
contact-range forces (the consensus seems to be that they
are not required at least up to
${\rm NLO}$~\cite{Birse:2009my,Song:2016ale,Zuo:2026uhi},
though it has not been determined yet at which order they do appear).

\section{Renormalization group and density dependence}
\label{sec:density-dependence}

The top-down EFT formulation of nuclear matter I have previously adopted
might also be characterized as {\it vacuum-derived}.
The starting point are the few-nucleon forces in the vacuum, which are
then evolved into the infrared to derive power
counting rules for the EOS.

Yet, a {\it medium-derived} EFT formulation is not only possible
but necessary for uncovering infrared corrections to the EOS
that are missed in the vacuum centered approach.
By medium-derived I refer to the idea of introducing a cutoff in the
G-matrix and then observing what type of counterterms or other
correcting terms are forced if I impose cutoff independence.
In the vacuum case --- that is, if one derives the EFT from the requirement of
a cutoff-independent T-matrix --- this approach simply reproduces
the derivative expansion of the contact-range interactions.
In contrast, cutoff independence in nuclear matter leads to the appearance of
new compensating terms explicitly depending on the density of the system.
This includes terms that scale with fractional powers of the Fermi momentum,
which have been traditionally required to reproduce the compressibility of
nuclear matter.

\subsection{Compensating terms in the vacuum}

If one considers first the vacuum, the starting point is the relation
\begin{eqnarray}
  T(E + i\epsilon)\, | \vec{p}\, \rangle = V | \Psi_{\vec{p}\,}^{+} \rangle \, ,
\end{eqnarray}
which connects the two-body free and full (outgoing) wave functions,
$| \vec{p}\, \rangle$ and  $| \Psi_{\vec{p}\,}^{+} \rangle$,
by means of the T-matrix and the potential.
For convenience it will be better to work with what is often referred to
as the K-matrix, which is related to the T-matrix by
\begin{eqnarray}
  \frac{1}{K(E)} = {\rm Re}\left[ \frac{1}{T(E)} \right] \, .
\end{eqnarray}
If one projects into S-wave, the relation between the free and full
wave function now reads
\begin{eqnarray}
  K(E)\, | p \, \rangle = V | \Psi_{p} \rangle \, ,
\end{eqnarray}
with
\begin{eqnarray}
  \langle r | p \rangle = \frac{\sin{(p r)}}{p r}
  \quad \mbox{and} \quad
  \langle r | \Psi_p \rangle = \frac{u_p(r)}{r} \, ,
\end{eqnarray}
where $u_p(r)$ is the S-wave reduced wave function, defined
with the asymptotic ($r \to \infty$) normalization
\begin{eqnarray}
  u_{p}(r) \to \frac{\sin{(p\,r + \delta(p))}}{p \cos{\delta(p)}} \, ,
\end{eqnarray}
with $\delta(p)$ the S-wave phase shift.
If one now considers the on-shell K-matrix:
\begin{eqnarray}
  K(p) = \langle p | K(E = \frac{p^2}{m_N}) | p \rangle =
  -\frac{4\pi}{m_N}\, \frac{\tan{\delta(p)}}{p} \, ,
  \nonumber \\
\end{eqnarray}
then it can be expressed as the r-space integral
\begin{eqnarray}
  K(p) = 4\pi \int_{0}^{\infty}\,dr\,V(r)\,\frac{\sin{(p r)}}{p}\,u_{p}(r) \, ,
\end{eqnarray}
where the potential and reduced wave functions are the exact ones, which
are usually unknown at short distances.

Now one may separate the integral in two parts by means of a cutoff
\begin{eqnarray}
  K(p) &=& 4\pi\,\int_{R_c}^{\infty}\,dr\,V(r)\,\frac{\sin{(p r)}}{p}\,u_{p}(r)
  \nonumber \\
  &+& 4\pi\,\int_{0}^{R_c}\,dr\,V(r)\,\frac{\sin{(p r)}}{p}\,u_{p}(r) \, ,
\end{eqnarray}
which might be used as a tool to isolate the unknown physics at
short distances.
Within the second integral ($r < R_c$), it is in principle
possible to expand the reduced incoming and outgoing
wave functions in powers of the momentum
\begin{eqnarray}
  \frac{\sin(p r)}{p} &=& \sum_{n=0}^\infty\,(-1)^n\,\frac{p^{2n} r^{2n+1}}{(2n+1)!}
  \, , \nonumber \\
  u_p(r) &=& \sum_{n=0}^{\infty} u_{2n}(r)\,p^{2n} \, ,
\end{eqnarray}
which leads to the expansion
\begin{eqnarray}
  4\pi\,\int_{0}^{R_c}\,dr\,V(r)\,\frac{\sin{(p r)}}{p}\,u_{p}(r) =
  \sum_{n=0}^{\infty} v_{2n}(R_c)\,p^{2n} \, , \nonumber \\
\end{eqnarray}
with $v_{2n}(R_c)$ basically representing moments of
the potential $V(r)$ at short distances.
Provided the potential is regular and has a finite range
this expansion converges at low momenta, in which case the full $K$-matrix
can be perfectly recovered from the expansion
\begin{eqnarray}
  K(p) &=& 4\pi\, \int_{R_c}^{\infty}\,dr\,V(r)\,\frac{\sin{(p r)}}{p}\,u_{p}(r)
  \nonumber \\
  &+& \sum_{n=0}^{\infty} v_{2n}(R_c)\, p^{2n} \, . 
\end{eqnarray}
This is expected to be equivalent to substituting the potential at
distances $r < R_c$ by a series of contact-range terms
\begin{eqnarray}
  V(r) \to \sum_{n=0}^{\infty} C_{2n}(R_c)\,\delta_{2n}(r; R_c) \, ,
\end{eqnarray}
where $\delta_{2n}(r; R_c)$ is a regularization of a Dirac-delta
involving $2n$ derivatives.
They might be spatial or time derivatives, depending on whether one chooses
momentum- or energy-dependent contact-range interactions, yet they
have a clear Lagrangian interpretation.

The bottom-line is that by integrating out the short distance degrees of freedom
one recovers the usual contact-range structure of non-relativistic EFTs.
It is worth noting though that this formalism is non-perturbative in nature
and thus not really suited to analyze power counting in general: it cleanly
uncovers the lowest order contact-range interaction and its scaling,
but if fails when applied to the higher order contacts,
which are perturbative by virtue of being subleading.
Yet, the present formalism is easily extended to the perturbative case.
These issues will not be explained here though,
but in Appendix \ref{app:counting}.

\subsection{Compensating terms in the medium}

With the G-matrix the argument is analogous,
except for a few differences.
The first is the relation~\cite{Day:1967zza}
\begin{eqnarray}
  G(E)\,| \vec{p}\,\vec{P}\, \rangle = V\,|\Psi_{\vec{p}\,,\vec{P}\,} \rangle \, ,
\end{eqnarray}
which connects the free and correlated (or in-medium) two-body wave functions,
where now the dependence in the total momentum $\vec{P}$ appears.
The second difference is that there are no $i \epsilon$ factors if the two
fermions are below the Fermi sea, which removes the intermediate step of
defining a K-matrix equivalent.
With this for S-waves one ends up with
\begin{eqnarray}
  G(p,P; k_F) = 4\pi \int_{0}^{\infty}\,dr\,V(r)\,\frac{\sin{(p r)}}{p}\,
  u_{p,P}(r; k_F) \, , \nonumber \\
\end{eqnarray}
where the reduced wave function is normalized as
\begin{eqnarray}
  u_{p,P}(r; k_F) \to \frac{\sin{(p r)}}{p} \, ,
\end{eqnarray}
for $r > R^{*}$, with $R^*$ the healing distance (which is inversely
proportional to $k_F$).
The expansion of the reduced in-medium wave function is now
\begin{eqnarray}
  u_{p,P}(r; k_F) &=& u_0(r) + P\,u_1(r) \nonumber \\
  &+& p^2\,u_{2A}(r) + P^2\,u_{2B}(r) + \dots
  \, ,
\end{eqnarray}
where the expansion includes powers of $P$.
This generates new terms in the expansion of the integral
\begin{eqnarray}
  && 4\pi \int_{0}^{R_c}\,dr\,V(r)\,\frac{\sin{(p r)}}{p}\,
  u_{p,P}(r; k_F) \nonumber \\
  && \qquad = v_0(R_c) + v_1(R_c)\,P \nonumber \\
  && \qquad + \, v_{2A}(R_c)\,p^2 + v_{2B}(R_c)\,P^2 +
  \dots \, , 
\end{eqnarray}
that don't (and can't) appear in the vacuum Lagrangian as contacts:
a two-body potential explicitly depending on $P$
breaks Galilean invariance.

By separating the EOS in contributions above and below the cutoff scale and
then explicitly considering the effect of these non-Lagrangian terms
on the energy per particle, one finds:
\begin{eqnarray}
  \frac{E}{A} &=& \frac{E}{A} {\Big |}_{r \geq R_c} + \frac{E}{A} {\Big |}_{r < R_c}
  \nonumber \\
  &=& \frac{E}{A} {\Big |}_{r \geq R_c} + \sum_{b} c_n(R_c) k_F^{3+b} \, ,
\end{eqnarray}
where it is apparent that the non-Lagrangian terms manifest as
density-dependence that can be directly included in the EOS
without resort for the intermediate step of assuming
the existence of density-dependent
contact-range potentials.
In fact, the new density-dependent terms often depend on powers of the Fermi
momentum that do not appear at tree-level for normal contact-range
interactions, which always involve an even number of derivatives.
This is also the reason why the values of the sum index $b$ are
not explicitly indicated in the formula above.

Yet, density-dependent contacts are a really convenient bookkeeping
device for reproducing the corrections to the EOS that originate
from integrating out the short-range physics.
This implies that these energy-dependent contact interactions are not actual
contacts though and do not have to be iterated:
provided they originate from terms that do not have a Lagrangian
interpretation, they only make sense at tree level.
The value of their couplings is still allowed to receive
subleading order corrections, though.
It is important to stress that energy-density dependent terms are still
connected to (and can be in principle calculated from) few-nucleon
forces, or more properly, BMF effects originating
from these few-nucleon forces.

In terms of the EFT formalism, one might write the contact-range potential
as the sum of two contributions, a {\it genuine} one and a
{\it pseudo-potential}:
\begin{eqnarray}
  V_C = V_C^{\rm gen} + V_C^{\rm pseudo} \, ,
\end{eqnarray}
where the genuine contribution might be either a momentum- or energy-expansion
\begin{eqnarray}
  V_C^{\rm gen} = \sum_{n=0}^{\infty} C_{2n}\,\left( \frac{p^{2n} + {p'}^{2n}}{2}
  \right) \quad \mbox{or} \quad
  \sum_{n=0}^{\infty} C_{2n}\,k^{2n} \, , \nonumber \\
\end{eqnarray}
while the pseudo-potential contribution is density-dependent
\begin{eqnarray}
  V_C^{\rm pseudo} = \sum_{b} \tilde{C}_{b} k_F^{b} \, ,
\end{eqnarray}
where the expansion index $b$ is left unspecified as it depends
on the details of the system.
For distinction purposes, the couplings corresponding to the pseudo-potentials
have been marked with a tilde.
It is important to stress that the pseudo-potential contribution
is not meant to be iterated, but only used at tree level instead.

This convention allows for a clear distinction between contributions
originating from integrating out the two-body forces (genuine) or
BMF effects (pseudo-potential).
Nonetheless it might still be in principle possible to use a density-dependent
representation for the genuine contact-range interaction for specific purposes.
Particularly when they might be convenient, as in the previous discussions
about the power counting of few-body forces in the EOS.

\subsection{Density-dependent terms}

Now I will consider the specific form of the in-medium reduced wave
function at short distances and its relation with the power
counting of density-dependent terms.
The easiest case is that of a reduced wave function that is regular at
the origin that might be schematically expanded as
\begin{eqnarray}
  && u_{p,P}(r; k_F) = r + b_1 \, P \, r^2 
    + b_{2A} \, p^2 r^3 + b_{2B} \, P^2 r^3 + \dots  \, ,
    \nonumber \\
\end{eqnarray}
where the normalization of the term proportional to $r$ (equal to one)
has been determined from the condition of converging to the free solution
above the healing distance.
In this case the ${\rm LO}$ interaction is a momentum- and energy-independent
contact-range potential, which is also density-independent.
This reproduces the vacuum power counting of a theory
in which the scattering length is small.

Notice that there are density-dependent corrections, but they are subleading.
In particular the BMF effects generating these density dependent
corrections are perturbative themselves.
As a consequence they are reproducible in terms of the iterations of
the usual (or genuine) EFT potential and no pseudo-potential
contribution is actually required.

The next case is when the reduced wave function is irregular at the origin
\begin{eqnarray}
  u_{p,P}(r; k_F) =
  \mathcal{N}(k_F)\,\left[ 1 + b_1 P r + \dots \right] \nonumber \, , \\
\end{eqnarray}
where now it is necessary to include a normalization factor $\mathcal{N}(k_F)$.
The scaling of the normalization factor with $k_F$ is easy to determine from
the condition that the wave function heals around $k_F r \sim 1$:
\begin{eqnarray}
  && u_{p,P}(r \sim \frac{1}{k_F}; k_F) \sim \frac{1}{p}\,\sin{(\frac{p}{k_F})}
  \nonumber \\
  && \qquad \qquad \sim \frac{1}{k_F} - \frac{p^3}{6\,k_F^2} +
  \mathcal{O}\left( \frac{p^5}{k_F^4} \right) \, ,
\end{eqnarray}
where in the second line one expands in powers of $p / k_F$.
This leads to
\begin{eqnarray}
  \mathcal{N}(k_F) \sim \frac{1}{k_F} \, ,
\end{eqnarray}
which implies the existence of a correction term that explicitly
depends on $k_F$, that is:
\begin{eqnarray}
    && 4\pi \int_{0}^{R_c}\,dr\,V(r)\,\frac{\sin{(p r)}}{p}\,
  u_{p,P}(r; k_F) \nonumber \\
  && \qquad = \frac{v_{-1}(R_c)}{k_F} + v_0(R_c) +  \dots \, . 
\end{eqnarray}
Equivalently, if one relies on density-dependent pseudo contact-range
interactions, the ${\rm LO}$ contact-range potential will
contain a genuine and pseudo-potential contribution
\begin{eqnarray}
  V^{\rm LO}_C = C_0 + \frac{\tilde{C}_{-1}}{k_F} \, .
\end{eqnarray}
With this contact the ${\rm LO}$ EOS reads:
\begin{eqnarray}
  \frac{E}{A} \Big|_{\rm LO} &=& t \rho^{2/3} + c_{-1} \rho^{2/3} + c_0\,\rho \, ,
\end{eqnarray}
where the $c_{-1}$ term originating from the density-dependent pseudo contact 
displays the same density-dependence as the kinetic term.
This is not arbitrary: this term is basically there because
the irregular wave function corresponds to expanding around
the unitary limit.
This in turn requires a contribution to reproduce the Bertsch
parameter~\cite{Baker:1999np}, which is a BMF effect as it
requires the full iteration of the LO vacuum contact.

\subsection{Fractional powers of the Fermi momentum}

Finally, I consider the case when the reduced wave function behaves as $r^a$
with the exponent $a$ not necessarily integer:
\begin{eqnarray}
  && u_{p,P}(r; k_F) = \nonumber \\ && \qquad
  \mathcal{N}(k_F)\,\left[ r^a f_0(Q r) + b_1 P r^{a+1} f_1(Q r) +
    \dots \right] \, , \nonumber \\
\end{eqnarray}
where $f_0(x)$, $f_1(x)$ are oscillatory functions of order one,
usually taking the form of a trigonometric (e.g. sine, cosine)
function with arguments $1/x^{2a-1}$ (and $Q$ representing
a light scale related to the strength of the potential
generating the wave function).
Matching to the free wave function at $k_F r \sim 1$,
the normalization factor of this wave function scales as
\begin{eqnarray}
  \mathcal{N}(k_F) \propto k_F^{a-1} \, ,
\end{eqnarray}
leading to a density-dependent correction of the type
\begin{eqnarray}
  V_C^{\rm pseudo} = \tilde{C}_{a-1}\,k_F^{a-1} \, .
\end{eqnarray}
This type of reduced wave function appears as the non-perturbative solution
of inverse power-law attractive singular potentials, where for
$V(r) \propto -1/r^n$ the exponent of the wave function
is $a = n/4$.
That is, one ends up with a density dependence involving a fractional
exponent of the Fermi momentum.

If the behavior of the wave function below the healing distance is driven
by subleading TPE ($n=6$), which is by far the strongest contribution to
the chiral potential at intermediate distances, the density dependent
term will be proportional to $k_F^{1/2}$ or $\rho^{1/6}$.
This coincides with one the most common exponents appearing
in density-dependent Skyrme forces.
In this case at ${\rm LO}$ the effective interaction (contact plus
pseudo contact) is
\begin{eqnarray}
  V^{\rm LO}_C = C_0 + \tilde{C}_{1/2}\,k_F^{1/2} \, ,
\end{eqnarray}
and the EOS reads
\begin{eqnarray}
  \frac{E}{A}\,\Big|_{\rm LO} = t\,\rho^{2/3} + c_0\,\rho + c_{1/2}\,\rho^{1+1/6} \, ,
\end{eqnarray}
where it is important to keep in mind that the fractional density dependence
does not come from a genuine contact-range potential, but is instead
the result of integrating out the BMF effects
coming from subleading TPE, which can be encoded
in a density-dependent pseudo-potential.

This is a really exciting result.
First, it provides a simple explanation for the $t_3$ density-dependent term
typically appearing in Skyrme forces:
\begin{eqnarray}
  \delta V_{\rm Skyrme} = \frac{1}{6}\,t_3\,(1 + x_3 P_{\sigma})\,\rho^{\alpha} \, ,
\end{eqnarray}
where $t_3$ is the coupling, $x_3$ a parameter and
$P_{\sigma} = (1 + \vec{\sigma}_1 \cdot \vec{\sigma}_2)/2$
the spin-exchange operator (with eigenvalues -1 and +1
for the singlet and triplet, respectively).
Second, it also explains one of its most usual exponents, $\alpha = 1/6$.
Third, by not associating the microscopic explanation of this term
with a three-body force, but with a isoscalar and central two-body
force --- which happens to be the strongest medium-range
($1 \leq m_{\pi}r \leq 2$, with $m_{\pi}$ the pion mass)
contribution to the EFT potential by far --- there is no mystery
as to why this term is usually included
in neutron matter calculations.

There are caveats as well: for starters, while the exponent in the Fermi
momentum of the pseudo contact depends explicitly on the behavior of
the wave function at distances that are formally long-range,
the truth is that the scale separation in nuclear EFT
is not particularly clean.
That is, the $1 \leq m_{\pi}r \leq 2$ distance range which determines
the fractional exponents are what could be denoted as {\it intermediate scales},
sitting uncomfortably in between the long- and short-range
scales of the two-nucleon system.
Thus, a certain degree of skepticism is justified regarding whether
the reduced wave function really displays the $r^{3/2}$ power-law
dependence required to explain the $\alpha = 1/6$ exponent.

Yet, predictions derived from subleading TPE alone tend to work well.
For instance, the singlet and triplet two-nucleon phase shifts can be
effectively described in terms of {\it chiral van der Waals
  forces}~\cite{PavonValderrama:2005wv},
which refer to  the chiral limit ($m_{\pi} \to 0$) of subleading TPE.

Besides, this explanation requires that it should be
possible to formulate an EFT in which subleading two-pion exchange
is a ${\rm LO}$ effect.
This idea was considered a long time ago~\cite{Valderrama:2010aw},
mostly as a pathology, where only recently has this alternative
EFT expansion been explicitly shown to be a viable
counting~\cite{Valderrama:2021bql} (a more intuitive explanation
of non-perturbative TPE in the singlet can be found
in~\cite{Mishra:2021luw}, though it does not explore
the EFT expansion resulting from this choice).
Indeed this expansion works perfectly well in two-nucleon scattering, 
outperforming more conservative EFT expansions in the singlet.
But it remains unclear whether non-perturbatively renormalized
subleading TPE can by itself generate saturation.
While traditionally saturation requires the inclusion of three-nucleon forces,
it is also true that renormalized OPE is able to saturate at cutoffs
$\Lambda \gtrsim 1\,{\rm GeV}$~\cite{Machleidt:2009bh}, though
the resulting saturation properties are far from ideal.
Thus, it would be really interesting to know what happens with
renormalized TPE and how much it improves over renormalized OPE.

\section{Discussion and conclusions}
\label{sec:discussion}

In this manuscript I have considered the EFT description of nuclear matter.
The observation underlying the present analysis is the special role of
the healing distance in the infrared RG evolution of
contact-range couplings in nuclear matter.
The idea is that as the two-nucleon wave function evolves into a free wave
function at interparticle separations larger than the healing distance,
the running of the contact-range couplings essentially freezes.
The resulting EFT description is thus different than in the vacuum: naively,
couplings that do not run are expected to be perturbative.
This intuition has been explicitly confirmed from analyzing the G-matrix and its
corresponding contribution to the energy per nucleon in a contact-range
theory, where loops happen to be suppressed in the medium.

The previous analysis relies however on knowing the running of the coupling
constants in nuclear matter at finite density, which is not always the case.
Yet, this is not a requirement and one might as well use the vacuum running
of the couplings at the cutoff $\Lambda^*$ at which they are expected to freeze
at a given density.
This results in a power counting in which the loop integrals in nuclear matter
are counted as $(k_F^* - k_F)$, instead of the usual $Q$, where $k_F^*$ is
proportional to $\Lambda^*$.
This is interesting for the following reasons:
\begin{itemize}
\item[(i)] Independently of whether one is dealing with vacuum or
  nuclear matter, loops formally scale as $Q$ 
  (because $k_F^* - k_F$ is after all a light scale).
\item[(ii)] Yet there is a density (of one's own choosing) for which loop
  corrections vanish, thus explaining why the ${\rm LO}$ couplings
  do not have to be iterated in nuclear matter.
  That is, they only enter at tree level in the ${\rm LO}$ EOS,
  at least in the vicinity of $k_F^*$, which is where
  the EFT expansion of the EOS is optimized.
\end{itemize}
Probably a good choice for $k_F^*$ is the Fermi momentum reproducing
the equilibrium density of symmetric nuclear matter.
Of course this expansion does not necessarily work far from $k_F^*$, yet the EOS
is by itself not an observable: what matters is either its
equilibrium properties (density, energy per nucleon, compressibility) or
the observable properties of the system for which
the EOS is an input (e.g. the maximum mass of
a neutron star, its tidal deformability, etc.).
Besides, once the initial expansion around a given $k_F^*$ fails, there is
always the possibility of redefining the EFT expansion of the EOS around
a second Fermi momentum $k_F^{*'}$.

The power counting of the contact-range couplings in nuclear matter is
in principle inherited from the vacuum.
Thus one is free to choose among the different countings available
in the few-nucleon system.
Yet, there is a caveat: if one wants loop contributions to be finite
then it is necessary to include auxiliary or redundant counterterms,
that is, counterterms that do not carry physical information
but are there to remove residual cutoff dependence (a more detailed
explanation may be found in~\cite{PavonValderrama:2026pmj}).
The reason is that the EFT expansion of nuclear matter is perturbative,
which implies that the matching with the RG evolution of
few-body forces happens at soft cutoffs.
If one expands around the soft cutoff limit, the (originally harmless
in the ultraviolet) residual cutoff dependence manifests as divergences.
If the auxiliary counterterms absorbing the residual cutoff dependence are
there in the first place, the divergences are removed.
But if the auxiliary counterterms are not present, which is what happens with
most EFT expansions in the few-body sector, then new counterterms have to be
introduced to cure the divergences.
In general, if this is the case, one is effectively forced to add new
parameters to the EFT description of nuclear matter.
The reason is that without the auxiliary counterterms there is no right
guess about the size of the finite part of the new counterterms.

Thus there are two approaches to the power counting of nuclear matter
\begin{itemize}
\item[(a)] A top-down approach, where power counting coincides with
  that of the vacuum at all orders, but which requires the inclusion of
  auxiliary counterterms not encoding physical information.
\item[(b)] A bottom-up approach, where power counting is eventually
  modified from the necessity of absorbing divergences that appear
  in many-body perturbation theory (even though these 
  divergences do vanish after resummation).
\end{itemize}
Both approaches are formally equivalent once all many-body perturbative
corrections are included, and hence their difference is practical
rather than theoretical.
But if one is interested in deriving the nuclear matter EOS directly from
few-nucleon forces, it will be necessary to extend the concept of
{\it improved actions}~\cite{Contessi:2023yoz,Contessi:2024vae,Contessi:2025xue}
(equivalent to the idea of auxiliary
counterterms~\cite{PavonValderrama:2026pmj})
from ${\rm LO}$ to subleading orders.

Besides the contact-range interactions already appearing in the EFT expansion
of the two-nucleon system, their infrared RG evolution at finite density
(which involves the analysis of the G-matrix) implies the existence of
new compensating or correcting terms that appear in nuclear matter.
These are density-dependent terms that might be directly included in the EOS
or indirectly by means of a contact-range pseudo-potential.
By a pseudo-potential it is meant a term formally identical to a contact-range
potential except for the fact that it should not be iterated
(as it does not represent a real two-body interaction).
This provides a clean explanation of the density-dependent terms
usually appearing in the Skyrme and Gogny
interactions~\cite{Skyrme:1959zz,Decharge:1979fa}.
In particular the RG evolution of chiral TPE in the infrared explains
one of the most typical exponents for the first density-dependent
term (the $t_3$ term) in Skyrme.

The breakdown scale of the EFT expansion of nuclear matter is in principle
identical to that of the two-nucleon sector, as previously
argued in the literature~\cite{Furnstahl:1999ix}.
The reason is that the T- and G-matrices share the same short-range behavior,
and thus lead to the same breakdown scale once the center-of-mass
momentum $k$ is substituted with the Fermi momentum $k_F$.
Usual estimations of the breakdown scale of two-nucleon scattering lie in
the $M \approx (400-600)\,{\rm MeV}$ range~\cite{Furnstahl:2015rha,Melendez:2017phj,Millican:2024yuz,Millican:2025sdp},
suggesting a {\it breakdown density} of
\begin{eqnarray}
  \rho_{\rm M} \approx (3-12)\,\rho_0 \, ,
\end{eqnarray}
with $\rho_0$ the equilibrium density of symmetric nuclear matter.
One might distinguish though between {\it intrinsic} and {\it extrinsic}
breakdown scales, where the first refers to the convergence of
the perturbative expansion of finite-range forces (the most evident
example being KSW~\cite{vanKolck:1998bw,Chen:1999tn} and
its ${\rm N^2LO}$ failure~\cite{Fleming:1999ee}),
while the second to the genuine limitations coming from
short-range physics outside the control of the EFT.
It might happen that the perturbative expansion of OPE, which is the limiting
factor for the convergence of KSW~\cite{Fleming:1999ee}, is more convergent at
finite densities that in the vacuum thus making KSW a viable
counting in nuclear matter.
Yet, this is but a conjecture at this point: elucidating this question will
require adapting current investigations about the perturbativeness of
OPE~\cite{PavonValderrama:2016lqn,Kaplan:2019znu} to the finite
density case.

The ${\rm LO}$ contact-range potential and pseudo-potential depend partly
on the choice of a power counting (pionless~\cite{vanKolck:1998bw,Chen:1999tn},
KSW~\cite{Kaplan:1998tg,Kaplan:1998we}, pionful~\cite{Nogga:2005hy,Birse:2005um,Valderrama:2009ei,Valderrama:2011mv,PavonValderrama:2025azr,Long:2011qx,Long:2011xw,Long:2012ve})
and partly on the choices of which BMF effects to integrate
out and include in the pseudo-potential.
For a KSW-type counting (S-wave contacts at ${\rm LO}$, subleading pions) and
a pseudo-potential derived from the two-pion exchange wave function,
the ${\rm LO}$ EOS is the one obtained with the $t_0$-$t_3$ subset
of the Skyrme interaction.
This coincides with the findings from the Orsay group~\cite{Yang:2013bhd,Yang:2016mvq,Yang:2017shi,Burrello:2020myg},
though there is a crucial difference in regards to the status of the density
dependent term, which in the present approach is a pseudo-potential.
In this regard it will be really interesting to recalculate the ${\rm NLO}$
EOS in the framework of the Orsay group with the pseudo-potential
reinterpretation of the $t_3$ term, which will simplify
the calculations.

If the starting point is pionful EFT instead, where OPE is leading and there
is at least one ${\rm LO}$ P-wave contact-range potential, the ${\rm LO}$
EOS reads instead:
\begin{eqnarray}
  \frac{E}{A} \Big|_{\rm LO} &=& t\,\rho^{2/3} +
  c_0\,\rho + c_{2}\,\rho^{1+2/3} + c_{1/2}\,\rho^{1+1/6} \nonumber \\
  &+& \frac{E}{A} \Big|_{\rm OPE} \, , 
\end{eqnarray}
with the origin of the $c_{2}$ coefficient being now the P-wave contribution and
where the OPE contribution in the second line (not written down explicitly here)
has been calculated for instance in~\cite{Kaiser:2002jz,Holt:2013fwa}.
(which also includes expressions for the iteration of OPE,
necessary at ${\rm NLO}$ and beyond).
Though this pionful EOS has a more complicated subleading structure than the
$t_0$-$t_3$ subset, it will be interesting to adapt the formalism
of~\cite{Yang:2013bhd,Yang:2016mvq,Yang:2017shi,Burrello:2020myg,Yang:2021akb}
to it and explore its ${\rm NLO}$ predictions.

To summarize, the infrared RG evolution of nuclear matter differs from that of
few-body forces in the vacuum owing to the appearance of a new long
distance scale, the healing scale.
This leads to its own low energy EFT expansion, in which forces that were
non-perturbative in the vacuum become perturbative at finite densities.
In principle the RG evolution of the couplings in nuclear matter requires
the couplings to explicitly depend on the Fermi momentum (and thus
on density).
Yet this requirement can be waived, in which case one ends up with
an expansion that is optimized at a particular density of one's own choice.
This expansion also involves the inclusion of density-dependent terms
in the EOS that might in turn be conveniently represented
by a density-dependent pseudo-potential.
The lowest-order description of nuclear matter coincides with certain
subsets of the Skyrme interaction already identified in the literature~\cite{Yang:2013bhd,Yang:2016mvq,Yang:2017shi,Burrello:2020myg,Yang:2021akb},
though it might contain pionic contributions not present in Skyrme
depending on the choice of a power counting.

\section*{Acknowledgments}

I would like to thank Ludovic Bonneau for discussions and
C.-J. (Jerry) Yang for discussions and a careful reading
of the manuscript.
This work is partly supported by the National Science Foundation of
China under Grant Nos. W2543006 and 1252200936.

\appendix

\section{Expansion of the G-matrix}
\label{app:g-matrix}

Here I consider the calculation of the G-matrix for a contact-range theory
at the lowest order and its expansion in the infrared.
This is necessary for a detailed understanding of the scaling
properties of the loop corrections in nuclear matter.

\subsection{G-matrix for a contact-range interaction}

In the vacuum the on-shell T-matrix for a momentum- and energy-independent
contact-range potential is 
\begin{eqnarray}
  T(p) = \frac{1}{\frac{1}{C_0(\Lambda)} - I_0(p; \Lambda)} \, ,
\end{eqnarray}
where $p = \sqrt{m_N\,E}$, with $E$ the center-of-mass energy and $I_0$
the loop function, which is given by
\begin{eqnarray}
  I_0(p; \Lambda) =
  \int_{\Lambda}\,\frac{d^3 \vec{q}}{(2\pi)^3}\,\frac{1}{E - \frac{q^2}{m_N}} \, ,
\end{eqnarray}
where the subscript $\Lambda$ indicates that the integral is regularized.
The exact form of $I_0$ depends on the choice of the regulator.
If one regularizes with PDS, one obtains
\begin{eqnarray}
  I_0^{\rm PDS}(p; \Lambda) = -
  \frac{m_N}{4 \pi}\,\left( \Lambda + i p \right) \, ,
\end{eqnarray}
for the $E + i \epsilon$ prescription, where $\Lambda$ is
the PDS regularization (or renormalization) scale.
From the condition of reproducing the scattering length $a_0$ (or, equivalently,
the on-shell T-matrix at $p = 0$), one obtains the well-known running
\begin{eqnarray}
  \frac{1}{C_0^{\rm PDS}(\Lambda)} =
  \frac{m_N}{4\pi}\,\left( \frac{1}{a_0} - \Lambda \right) \, .
  \label{eq:C0-PDS}
\end{eqnarray}

In the medium the G-matrix for a contact-range interaction is structurally
identical to the T-matrix in the vacuum, that is:
\begin{eqnarray}
  G(p,P, k_F; \Lambda) =
  \frac{1}{\frac{1}{C_0(\Lambda)} - I_{0F}(p, P, k_F; \Lambda)}
  \, ,
  \label{eq:G-vacuum-condition}
\end{eqnarray}
where the most obvious difference is that the loop integral now depends
not only on the relative momentum $p$, but also
on the total and Fermi momenta, $P$ and $k_F$.
The loop integral reads
\begin{eqnarray}
  && I_{0F}(p, P, k_F; \Lambda) =
  \int_{\Lambda} \frac{d^3 q}{(2 \pi)^3}\,\frac{Q_F}{E - \frac{q^2}{m_N}}
  \nonumber \\
  && \qquad =
  \int_{\Lambda} \frac{d^3 q}{(2 \pi)^3}\,
  \frac{\theta(q_1 - k_F)\,\theta(q_2-k_F)}{E - \frac{q^2}{m_N}} \, ,
\end{eqnarray}
with the subscript $\Lambda$ again indicating regularization, and
where in the second line the form of the projector $Q_F$
is written explicitly, with
\begin{eqnarray}
  \vec{q}_1 = \frac{\vec{P}}{2} + \vec{q} \quad \mbox{and} \quad
  \vec{q}_2 = \frac{\vec{P}}{2} - \vec{q} \, ,
\end{eqnarray}
the momentum of each of the two fermions in the loop.
If one chooses a regularization that depends only on the relative momentum $q$,
then the angular part of the integral is independent of the specific details
of this regulator, in which case the loop integral can be written as
\begin{eqnarray}
  && I_{0F}(p,P,k_F; \Lambda)   \nonumber \\
  && \quad = \frac{m_N}{2\pi^2} \Big[
    \int_{\sqrt{k_F^2 - P^2/4}}^{k_F + {P}/2}\,dq\,
    \left( \frac{P^2/4 + q^2 - k_F^2}{P q} \right)
    \nonumber \\
    && \qquad \quad
    + \int_{k_F + P/2}^{\infty}\,dq\, \Big] \frac{q^2}{p^2 - q^2}\,
  g(\frac{q}{\Lambda}) \, ,
  \label{eq:I0F-integration}
\end{eqnarray}
with $g(x)$ the regulator function.
For a typical separable regulator, that is:
\begin{eqnarray}
  \langle p' | V_C | p \rangle = C_0(\Lambda)\,
  f(\frac{p'}{\Lambda})\,f(\frac{p}{\Lambda}) \, ,
\end{eqnarray}
one has $g(x) = f^2(x)$.
Besides, if the two particles are below the Fermi sea there is no need for
the $i \epsilon$ prescription and the integral is real.
 
In PDS the loop integral can be divided into two parts, one that depends on
the regularization scale $\Lambda$ and one that does not:
\begin{eqnarray}
  I_{0F}^{\rm PDS}(p,P,k_F;\Lambda)
  = - \frac{m_N}{4\pi}\,\Lambda + I_{0F}^{\rm reg}(p,P,k_F) \, ,
  \nonumber \\
    \label{eq:I0F-PDS-decomp}
\end{eqnarray}
where the scale-independent (or regular) piece can be further decomposed into
\begin{eqnarray}
  I_{0F}^{\rm reg}(p,P,k_F) = I_{0Fa}(p,P,k_F) + I_{0Fb}(p,P,k_F) \, ,
  \nonumber \\
  \label{eq:I0F-finite-PDS-decomp}
\end{eqnarray}
with
\begin{eqnarray}
  && I_{0Fa}(p, P, k_F) =
  - \frac{m_N}{8 \pi^2 P}\,\Big[ P\,(2 k_F + P) + \nonumber \\ &&
    \qquad  2\,(k_F^2 - {P^2}/{4} - p^2)\,\log{\left( \frac{k_F^2 - P^2/4 - p^2}{(k_F + P/2)^2 - p^2} \right)} \Big] \, , \nonumber \\
  \label{eq:I0F-comp-a} \\
  && I_{0Fb}(p, P, k_F) = 
  \frac{m_N}{4\pi^2}\,(2 k_F + P)\,\Big[ \nonumber \\
    && \qquad 1 - \frac{p}{k_F + P/2}\,{\rm atanh}\,
    \left( \frac{p}{k_F + P/2}\right)
  \Big] \, , \label{eq:I0F-comp-b} 
\end{eqnarray}
where they correspond to the two integration brackets
in Eq.~(\ref{eq:I0F-integration}), that is,
from $q = \sqrt{k_F^2 - P^2/2}$ to $k_F + P/2$ for $I_{0Fa}$ and
from $q = k_F + P/2$ to infinity for $I_{0Fb}$.

With the previously defined scale-independent part of the loop integral,
$I_{0F}^{\rm reg}$, the calculation of the G-matrix in PDS reads
\begin{eqnarray}
  G(p,P, k_F) = \frac{4\pi}{m_N}\,\frac{1}{\frac{1}{a_0} -
    \frac{4\pi}{m_N}\,I_{0F}^{\rm reg}} \, ,
  \label{eq:G-matrix-contact}
\end{eqnarray}
where the running of $C_0(\Lambda)$ has been taken from the vacuum case,
Eq.~(\ref{eq:C0-PDS}).

For a separable regulator the form of the loop integral 
follows the pattern:
\begin{eqnarray}
  I_{0F} = - \frac{m_N}{4\pi}\,c_{r}\,\Lambda + I_{0F}^{\rm reg} +
  \delta I_{0F}^{\rm res} \, , 
  \label{eq:I0F-decomp-arbitrary}
\end{eqnarray}
where $c_{r}$ is a regulator-dependent number, $I_{0F}^{\rm reg}$ the regular
(plus cutoff independent) part of the loop integral, and
$\delta I_{0F}^{\rm res}$ a contribution only
containing the residual cutoff dependence.
The constant $c_r$ is given by
\begin{eqnarray}
  c_r = \frac{2}{\pi}\,\,\int_0^{\infty} f^2(x)\,d x \, ,
\end{eqnarray}
where for a sharp-cutoff --- that is, $f(x) = \theta(1-x)$ ---
this numerical factor is $c_{r} = 2/\pi$.
The part of the loop integral encoding the residual cutoff dependence
can be written as
\begin{eqnarray}
  && \delta I_{0F}^{\rm res}(p,P,k_F; \Lambda)   \nonumber \\
  && \quad = \frac{m_N}{2\pi^2} \Big[
    \int_{\sqrt{k_F^2 - P^2/4}}^{k_F + {P}/2}\,dq\,
    \left( \frac{P^2/4 + q^2 - k_F^2}{P q} \right)
    \frac{q^2}{p^2 - q^2}\,
    \nonumber \\
    && \qquad \quad
    + \int_{k_F + P/2}^{\infty}\,dq\,\frac{p^2}{p^2-q^2}
    \nonumber \\
    && \qquad \quad
    + \int_{0}^{k_F + P/2}\,dq \Big]\,
  \left( f^2(\frac{q}{\Lambda}) - 1\right) \, ,
  \label{eq:I0F-residual}
\end{eqnarray}
and, while simple in the $\Lambda \gg k_F$ limit (where it scales as
$\mathcal{O}(\Lambda^{-1})$), it might become
really involved for $\Lambda \lesssim k_F$.
Note that the regular part of the loop integral has been defined as to coincide
with its evaluation in PDS, and its expression is identical to that
of $\delta I_{0F}^{\rm res}$ above but with the substitution
$( f^2(\frac{q}{\Lambda}) - 1) \to 1$.

With the previous components, the G-matrix for a separable regulator reads
\begin{eqnarray}
  G(p,P, k_F; \Lambda) = \frac{4\pi}{m_N}\,\frac{1}{\frac{1}{a_0} -
    \frac{4\pi}{m_N}\,\left( I_{0F}^{\rm reg} +
    \delta I_{0F}^{\rm res} \right)
  } \, , \nonumber \\
  \label{eq:G-matrix-contact-separable}
\end{eqnarray}
where the $C_0(\Lambda)$ from which it has been calculated is given by
\begin{eqnarray}
  \frac{1}{C_0(\Lambda)} = \frac{m_N}{4 \pi}\,
  \left( \frac{1}{a_0} - c_r\,\Lambda \right) \, .
\end{eqnarray}
Independently of the regulator choice, the $\Lambda \to \infty$ limit of
this G-matrix reproduces the result of Eq.~(\ref{eq:G-matrix-contact}).

\subsection{The G-matrix with a delta-shell regulator}

The calculation of the G-matrix in r-space is not entirely trivial.
The reason lies in the required set of techniques, which is very
different from the more familiar methods used in the p-space case. 

In particular I will derive the G-matrix from the form of the in-medium
wave function, where the starting point is the (S-wave)
Bethe-Goldstone equation
\begin{eqnarray}
  \Psi_{p,P}(r) = j_0(p\,r) +
  \int d^3 \vec{r}\,'\,K_0(|\vec{r} - \vec{r}\,'|)\,V(\vec{r}\,')\,
  \Psi_{p,P}(r') \, , \nonumber \\
\end{eqnarray}
with $\Psi_{p,P}$ the wave function, $j_0(x) = \sin{x}/x$ the spherical Bessel
function, and with the in-medium propagator $K_0$ given by
\begin{eqnarray}
  K_0(r) = \int \frac{d^3\vec{q}}{(2\pi)^3}\,
  \frac{m_N}{p^2 - q^2}\,e^{i \vec{q} \cdot \vec{r}}\,\theta(q_1 - k_F)\,\theta(q_2 - k_F) \, , \nonumber \\
\end{eqnarray}
which can be rewritten as
\begin{eqnarray}
  && K_0(r) = \frac{m_N}{2\pi^2} \Big[
    \int_{\sqrt{k_F^2 - P^2/4}}^{k_F + {P}/2}\,dq\,
    \left( \frac{P^2/4 + q^2 - k_F^2}{P q} \right)
    \nonumber \\
    && \qquad \quad
    + \int_{k_F + P/2}^{\infty}\,dq\, \Big] \frac{q^2}{p^2 - q^2}\,
  \frac{\sin{(q r)}}{q r} \, .
\end{eqnarray}
It is worth noting that there is a closed form solution for $K_0(r)$,
though it is rather unwieldy and will not be written down here.

If one regularizes the contact-range potential with a delta-shell
\begin{eqnarray}
  V_C(\vec{r}\,) = C_0(R_c)\,\frac{\delta(r- R_c)}{4\pi\,R_c^2} \, ,
  \label{eq:VC-delta-shell}
\end{eqnarray}
then it is useful to define the regularized in-medium propagator as
\begin{eqnarray}
  K_0(r; R_c) = \int_0^{\infty} dx\,\delta(x-R_c)\,\int
  \frac{d^2\hat{x}}{4 \pi}\,K(|\vec{r} - \vec{x}|) \, . \nonumber \\
\end{eqnarray}
For $r=R_c$ the calculation simplifies to
\begin{eqnarray}
  K_0(R_c; R_c) &=& 
  \int_{-1}^{+1} \,\frac{du}{2}\,K_0(R_c \sqrt{2-2u}) \nonumber \\
  &=& \frac{1}{2}\,\int_0^{2}\,y\,dy\,K_0(R_c \, y) \, ,
\end{eqnarray}
with the second line (which is the result of a change of variable) being
particularly easy to calculate.

In the $R_c \to 0$ limit the evaluation of $K_0(R_c; R_c)$ complies
with the expectation of a linear divergence plus a regular
contribution that already appeared in Eq.~(\ref{eq:I0F-PDS-decomp})
for the p-space calculation:
\begin{eqnarray}
  K_0(R_c; R_c) = - \frac{m_N}{4\pi}\,\frac{1}{R_c} + I_{0F}^{\rm reg}
  + \mathcal{O}(R_c) \, , \nonumber \\
\end{eqnarray}
with $I_{0F}^{\rm reg} = I_{0Fa} + I_{0Fb}$ as written
in Eqs.~(\ref{eq:I0F-finite-PDS-decomp}-\ref{eq:I0F-comp-b}),
from which it is apparent that $K_0(R_c; R_c)$ is simply
the loop function in r-space.

For finding the G-matrix, one first notices that the Bethe-Goldstone
equation for a delta-shell can be rewritten as
\begin{eqnarray}
  \Psi_{p,P}(r) &=& j_0(p\,r) + C_0(R_c)\,\Psi_{p,P}(R_c)\,K_0(r; R_c)
  \nonumber \\
  &=& j_0(p\,r) +
  \frac{j_0(p\,R_c)\,K_0(r; R_c)}{\frac{1}{C_0(R_c)} - K_0(R_c; R_c)} \, ,
\end{eqnarray}
where the denominator in the second line seems to be the G-matrix.
This identification has to be proven though, for which one might consider
Eq.~(\ref{eq:Psi-G-matrix}), which after taking matrix elements reads
\begin{eqnarray}
  \Psi_{p,P}(r) &=& j_0(p\,r) \nonumber \\
  &+& \int \frac{d^2\hat{r}}{4\pi}\,\int \frac{d^2\hat{p}}{4\pi}\,
  \langle \vec{r}\, | \frac{Q_F}{E-H_0} G(E) | \vec{p}\, \rangle
  \, , \nonumber \\
\end{eqnarray}
where the angular integrals are there to ensure
the projection into the S-wave.
Then, if one takes into account that in p-space the delta-shell potential
of Eq.~(\ref{eq:VC-delta-shell}) is separable:
\begin{eqnarray}
  \langle p' | V_C | p \rangle = j_0(p' R_c)\,C_0(R_c)\, j_0(p R_c) \, ,
\end{eqnarray}
then the matrix elements of the G-matrix are 
\begin{eqnarray}
  \langle p' | G(E) | p \rangle = j_0(p' R_c)\,G(E; R_c)\, j_0(p R_c) \, ,
  \label{eq:G-matrix-delta-shell}
\end{eqnarray}
where I have used the same symbol G for the operator and the part of
its evaluation that does not depend on the form factor (they are
easily distinguishable by the context).
With this, it follows that
\begin{eqnarray}
  &&
  \int \frac{d^2\hat{r}}{4\pi}\,\int \frac{d^2\hat{p}}{4\pi}\,
  \langle \vec{r}\, | \frac{Q_F}{E-H_0} G(E) | \vec{p}\, \rangle
  \nonumber \\ && \quad
  = 
  G(E; R_c)\,j_0(p R_c)\,\int \frac{d^3 \vec{q}}{(2\pi)^3}\,\frac{Q_F}{E - \frac{{q}^2}{m_N}}\,e^{i \vec{q} \cdot \vec{r}}\,j_0(q R_c) \, , \nonumber \\
\end{eqnarray}
which is not very revealing by itself until one uses the relation
\begin{eqnarray}
  j_0(q R_c) = \int d^3\,\vec{x}\,e^{- i \vec{q} \cdot \vec{x}}\,
  \frac{\delta(x-R_c)}{4\pi R_c^2} \, ,
\end{eqnarray}
from which one arrives at
\begin{eqnarray}
  &&
    \int \frac{d^2\hat{r}}{4\pi}\,\int \frac{d^2\hat{p}}{4\pi}\,
    \langle \vec{r}\, | \frac{Q_F}{E-H_0} G(E) | \vec{p}\, \rangle
    \nonumber \\ && \quad
    = 
    G(E; R_c)\,j_0(p R_c)\,K_0(r; R_c) \, ,
\end{eqnarray}
or, equivalently:
\begin{eqnarray}
  \Psi_{p,P}(r) = j_0(p\,r) + j_0(p\,R_c)\,G(E; R_c)\,K_0(r; R_c) \, .
  \nonumber \\
\end{eqnarray}
From this, the G-matrix ends up being
\begin{eqnarray}
  G(E; R_c) = \frac{1}{\frac{1}{C_0(R_c)} - K_0(R_c; R_c)} \, ,
\end{eqnarray}
with its $R_c \to 0$ limit  given by Eq.~(\ref{eq:G-matrix-contact}).

\subsection{In-medium renormalization of the G-matrix}

At zero external momenta ($p = P = 0$) the G-matrix reads
\begin{eqnarray}
  G(0,0,k_F) =
  \frac{4\pi}{m_N}\,\frac{1}{\frac{1}{a_0} - \frac{2}{\pi}\,k_F} \, .
\end{eqnarray}
This limit provides a different renormalization condition
for the contact-range coupling.
In this case the G-matrix is calculated as usual
\begin{eqnarray}
  G(p,P,k_F; \Lambda) =
  \frac{1}{\frac{1}{C_{0F}(\Lambda)} - I_{0F}(p,P,k_F;\Lambda)}
  \, , \nonumber \\
\end{eqnarray}
but with $C_{0F}(\Lambda)$ determined from the condition of
reproducing $G(0,0,k_F)$, which is why $C_{0F}$ is
referred to as the in-medium coupling.
That is, one is not required to calibrate the coupling to
the $p=0$ limit of the T-matrix.
Yet, both renormalization conditions
lead to the same G-matrix in the $\Lambda \to \infty$ ($R_c \to 0$) limit.
The difference with the in vacuum renormalization
condition of Eq.~(\ref{eq:G-vacuum-condition})
lies in the fact that its in medium version is able to handle cutoffs
that are softer than the Fermi momentum $k_F$.

For a sharp cutoff in momentum space one has
\begin{eqnarray}
  && I_{0F}(0,0,k_F;\Lambda) = -\frac{m_N}{2 \pi^2}\,\int_{k_F}^{\Lambda}\,dq
  \nonumber \\
  && \qquad = -\frac{m_N}{2 \pi^2}\,(\Lambda - k_F)\,\theta(\Lambda - k_F) \, ,
\end{eqnarray}
and by setting the coupling to reproduce $G(0,0,k_F)$ one obtains
the $C_{0F}(\Lambda)$ of Eq.~(\ref{eq:C0F-sharp-cutoff}).

In r-space at $p = P = 0$ the propagator $K_0$ reads
\begin{eqnarray}
  K_0(r) &=& -\frac{m_N}{2\pi^2}\,\int_{k_F}^{\infty} \frac{\sin{(q r)}}{q r}\,dr
  \nonumber \\
  &=& -\frac{m_N}{2\pi^2}\,\frac{1}{r}\,\left( \frac{\pi}{2} - {\rm Si}(k_F r)
  \right) \, ,
\end{eqnarray}
which for a delta-shell regulator leads to the $K_0(r;R_c)$
of Eq.~(\ref{eq:K0-delta-shell}), from which the wave
function in Eq.~(\ref{eq:wf-in-medium-delta-shell})
follows and thus the running of the $C_{0F}(R_c)$
coupling of Eq.~(\ref{eq:C0F-delta-shell}).

Finally, for PDS there is no difference for the coupling independently of
whether the renormalization condition is given by the $p=0$ T-matrix or
the $p=P=0$ G-matrix.
The reason is that PDS regularization is not sensitive to the existence of
the Fermi momentum as a low energy scale that impacts
the behavior of loops.

\subsection{The G-matrix in the infrared}
\label{subapp:G-infrared}

If one now rewrites the G-matrix as
\begin{eqnarray}
  && G(p,P, k_F) =  \nonumber \\ && \quad
  \frac{4\pi}{m_N}\,\frac{1}{\frac{1}{a_0} - \frac{2}{\pi} k_F -
    \frac{4\pi}{m_N}\,\left( I_{0F}^{\rm reg} - \frac{m_N}{2\pi^2} k_F\right)}
  \, , \nonumber \\
  \label{eq:G-matrix-infrared}
\end{eqnarray}
then, at low external momenta, it is possible to expand it
in inverse powers of $(\frac{1}{a_0} - \frac{2}{\pi}\,k_F)$
\begin{eqnarray}
  && G(p,P, k_F) =  
  \frac{4\pi}{m_N}\,\frac{1}{\frac{1}{a_0} - \frac{2}{\pi} k_F}\,
  \nonumber \\ && \quad \times
  \sum_{n=0}^{\infty} {\left[ \frac{4\pi}{m_N}\,\frac{1}{\frac{1}{a_0} - \frac{2}{\pi} k_F}\,\left( I_{0F}^{\rm reg} - \frac{m_N}{2\pi^2} k_F\right) \right]}^n
  \nonumber \\ && \quad
  = \sum_{n=0}^{\infty}\,G^{[n]}(p,P,k_F) \, ,
\end{eqnarray}
where $G^{[n]}$ refers to the $n$-th term contribution in this expansion,
which in turn can be rewritten as
\begin{eqnarray}
  && G^{[n]}(p,P,k_F) = \nonumber \\
  && \qquad G^{[0]}(k_F)\,{\left[ G^{[0]}(k_F) \left( I_{0F}^{\rm reg} - \frac{m_N}{2\pi^2} k_F\right) \right]}^n \, .
  \nonumber \\
\end{eqnarray}
By averaging the expansion of the G-matrix over states in the Fermi sea,
one obtains the corresponding expansion for the energy per nucleon
\begin{eqnarray}
  \frac{E}{A} = \sum_{n=0}^{\infty}\,\frac{E^{[n]}}{A} \, .
\end{eqnarray}
Then a direct calculation shows:
\begin{eqnarray}
  \frac{E^{[1]}}{E^{[0]}}
  &=& -\frac{k_F}{\frac{1}{a_0} - \frac{2}{\pi} k_F}\,
  \left( \frac{4 + 12 \log{2}}{35\pi} \right) \nonumber \\
  &=& -\frac{k_F}{\frac{1}{a_0} - \frac{2}{\pi} k_F}\,h_F(0)
  \, ,
\end{eqnarray}
where in the second line the numerical factor is written in terms of
the function $h_F(x)$ that appeared in Eq.~(\ref{eq:hF-ratio}).

This expansion can also be mapped into the perturbative expansion of
the G-matrix in powers of the in-medium coupling $C_{0F}$:
\begin{eqnarray}
  G = C_{0F}(\Lambda) + C_{0F}^2(\Lambda) I_{0F}(p,P,k_F;\Lambda) + \dots \, ,
  \nonumber \\
\end{eqnarray}
though the terms of this expansion are not cutoff independent.
For achieving a renormalized expansion one has to expand first the coupling
$C_{0F}$ in inverse powers of $(\frac{1}{a_0} - \frac{2}{\pi}\,k_F)$.
For cutoff regularization this yields
\begin{eqnarray}
  \frac{1}{C_{0F}(\Lambda)} &=& \frac{m_N}{4\pi}\,\left(
  \frac{1}{a_0} - \frac{2}{\pi}\,k_F - \frac{4\pi}{m_N} I_{0F}(0,0,k_F;\Lambda)
  \right) \nonumber \\
  &=& \frac{1}{C_{0F}^{[0]}} - I_{0F}(0,0,k_F;\Lambda) \, ,
\end{eqnarray}
from which one obtains the infrared expansion:
\begin{eqnarray}
  && G(p,P,k_F; \Lambda) = C_{0F}^{[0]} \nonumber \\
  && \qquad \Big[ 1 + C_{0F}^{[0]}\,\left(
    I_{0F}(p,P,k_F;\Lambda) - I_{0F}(0,0,k_F;\Lambda) \right)
    \nonumber \\
    && \qquad \quad \, + \dots \Big] \, ,
\end{eqnarray}
where the dots indicate the continuation of the geometric expansion.
In this case (i.e. with a finite cutoff), the relative size of the first
loop correction is
\begin{eqnarray}
  \frac{E^{[1]}}{E^{[0]}}
  &=& -\frac{k_F}{\frac{1}{a_0} - \frac{2}{\pi} k_F}\,h_F(\frac{k_F}{\Lambda})
  \, ,
\end{eqnarray}
with the function $h_F$ defined as 
\begin{eqnarray}
  && h_F(\frac{k_F}{\Lambda}) = 
  - \frac{1}{k_F}\,{\left( \frac{6 \pi^2}{k_F^3} \right)}^2\, \nonumber \\ && \quad \times
  \frac{4\pi}{m_N}\,\int_{k_1, k_2 < k_F} \frac{d^3\,\vec{k}_1}{(2\pi)^3}\,
   \frac{d^3\,\vec{k}_2}{(2\pi)^3} \nonumber \\
  && \qquad  \Big[
     I_{0F}(p,P,k_F;\Lambda) - I_{0F}(0,0,k_F;\Lambda) \Big] \, ,
   \nonumber \\
\end{eqnarray}
where by making suitable changes of variables it can be shown
that $h_F$ does indeed only depend on the ratio $k_F / \Lambda$.

In general the closed-form evaluation of this integral is not possible
except in a few particular cases.
For instance, for a sharp cutoff regulator and a cutoff $\Lambda > 2 k_F$
one has
\begin{eqnarray}
  && h_F^{\rm SC} (x) = - \frac{2}{35\,\pi\,x^7}\,\Big[ 30\,x^2 + 99\,x^4 + 17\,x^6
    -2 x^7 \nonumber \\ && \qquad
    + 6 x^3\,\left( 3 x^4 - 35 \right)\,{\rm atanh}(x) -6 x^7\,\log{2}
    \nonumber \\ && \qquad
    + (30 - 126\,x^2)\,\log{(1-x^2)}\Big] \quad
  \mbox{(for $x < 1/2$)} \, , \nonumber \\
  \label{eq:hF-SC-I}
\end{eqnarray}
while for softer cutoffs the form of $h_F^{\rm SC}$ becomes
considerably more involved.

It is not difficult to check that the $\Lambda \to \infty$ limit of
$h_F$ is regulator-independent for separable regulators.
This follows from the relation
\begin{eqnarray}
  && \lim_{\Lambda \to \infty}\,
  \left( I_{0F}(p,P,k_F;\Lambda) - I_{0F}(0,0,k_F;\Lambda) \right) 
  \nonumber \\
  && \qquad =
  I_{0F}^{\rm reg}(p,P,k_F) - \frac{m_N}{2\pi^2}\,k_F   \, ,
  \label{eq:I0F-diff}
\end{eqnarray}
which is a consequence of Eq.~(\ref{eq:I0F-decomp-arbitrary}) and
\begin{eqnarray}
  && I_{0F}(0,0,k_F;\Lambda) =
  -\frac{m_N}{2\pi^2}\,\int_{k_F}^{\infty}\,f^2(\frac{q}{\Lambda})\,dq
  \nonumber \\ && \qquad =
  \frac{m_N}{2\pi^2}\,\int_{0}^{k_F}\,f^2(\frac{q}{\Lambda})\,dq
  -\frac{m_N}{4\pi}\,c_r\,\Lambda 
  \, . 
\end{eqnarray}
Then by averaging over the states within the Fermi sea
$h_F(0)$ is recovered.

\section{Loop integrals in the medium}
\label{app:loop}

Here I briefly explain how to calculate the loop integrals
that are required for the order-by-order renormalization
of the EOS, as discussed in Section~\ref{sec:nuclear-matter}.
The evaluation of these integrals is necessary for deriving 
the $k_F^* - k_F$ scaling on which the power counting
for nuclear matter is based.

For a contact-range theory at tree level the two-body potential energy
per nucleon is given by
\begin{eqnarray}
  \frac{E_2}{A} &=& \frac{1}{2}\,\left( 1 - \frac{1}{g} \right)\,\rho\,
  \nonumber \\
  &\times&
  \left[ C_0 + \kappa_2\,C_2\,k_F^2
    + \kappa_4\,C_4\,k_F^4 + \dots \right] \, ,
  \nonumber \\
\end{eqnarray}
with $\kappa_{2n}$ a number that depends on the choice of representation
of the contact-range potential.
For density-dependent couplings this numerical factor is always equal to one
($\kappa_{2n} = 1$),
while for energy- and momentum-dependent couplings it is
\begin{eqnarray}
  \kappa_{2n} = \frac{18}{(n+2) (n+3) (2n+3)} \, . \label{eq:kappa-2n}
\end{eqnarray}
The first loop correction can be schematically written as
\begin{eqnarray}
  && \frac{\delta E_2}{A} = \frac{1}{2}\,\left( 1 - \frac{1}{g} \right)\,\rho\,
  \nonumber \\
  && \quad \times \, {\left( \frac{g}{\rho} \right)}^2\,
  \Big[ C_0^2\,I_{0F}(k_F; \Lambda) +
    2\,C_0 C_2\,I_{2F}(k_F; \Lambda) +
    \nonumber \\ && \qquad \qquad \quad
    C_2^2\, I^{(a)}_{4F}(k_F; \Lambda) + 2\,C_0 C_4\,I^{(b)}_{4F}(k_F; \Lambda)
        + \dots \Big] \, ,
  \nonumber \\
\end{eqnarray}
where the evaluation of the loop integrals depends on the chosen
regularization method and representation of
the contact-range potential. 
In what follows the integrals will be regularized in PDS,
which generates relatively simple expressions.

\subsection{Lowest order contact}

The loop integral corresponding to the first iteration of $C_0$ is relatively
easy to calculate: one begins with $I_{0F}$ as originally calculated
in Eqs.~(\ref{eq:I0F-PDS-decomp}-\ref{eq:I0F-comp-b}) and
averages it for states within the Fermi sea
\begin{eqnarray}
  && I_{0F}(k_F; \Lambda) = \nonumber \\
  && \quad \int_{k_1 < k_F} \frac{d^3\,\vec{k}_1}{(2\pi)^3}\,
  \int_{k_2 < k_F} \frac{d^3\,\vec{k}_2}{(2\pi)^3}\,I_{0F}(p,P,k_F;\Lambda) \, ,
  \nonumber \\
\end{eqnarray}
where the notation $I_{0F}$ refers either to the loop integral or its average,
with the distinction lying in the variables on which it depends.

For its evaluation,
one performs a change of variables from the momenta of
particles $1$ and $2$ to the total and relative momentum of
the two particle system. That is, the integral for states within
the Fermi sea is rewritten as
\begin{eqnarray}
  && \int_{k_1 < k_F} \frac{d^3\,\vec{k}_1}{(2\pi)^3}\,
  \int_{k_2 < k_F} \frac{d^3\,\vec{k}_2}{(2\pi)^3}
  = \nonumber \\
  && \qquad \int_0^{2 k_F} \frac{P^2 dP}{2\,\pi^2}\,
  \int_0^{\sqrt{k_F^2 - P^2/4}} \frac{p^2 dp}{2\,\pi^2}\, \times
  \nonumber \\ && \qquad \qquad
  \int \frac{d^2\hat{p}}{4\,\pi}\,
  \theta(k_F - k_1)\,\theta(k_F - k_2) \, .
  \nonumber \\
\end{eqnarray}
After doing the angular part, the loop integral further simplifies to
\begin{eqnarray}
  && I_{0F}(k_F; \Lambda) = \nonumber \\
  && \quad \int_0^{2 k_F} \frac{P^2 dP}{2\pi^2} \Big[
    \int_0^{k_F - P/2}\,\frac{p^2 dp}{2\,\pi^2} + \nonumber \\ && \qquad
    \int_{k_F-P/2}^{\sqrt{k_F^2 - P^2/4}}\,\frac{p^2 dp}{2\,\pi^2}\,
    \frac{k_F^2 - (P^2/4 + p^2)}{P p}
    \Big] \times \nonumber \\ && \qquad \qquad
  I_{0F}(p,P,k_F; \Lambda) \, .
\end{eqnarray}
Following Eqs.~(\ref{eq:I0F-PDS-decomp}) and (\ref{eq:I0F-finite-PDS-decomp}),
the integral in PDS can be separated into a diverging and
regular component
\begin{eqnarray}
  && I_{0F}(k_F; \Lambda) = -\frac{1}{{(2\pi^2)}^2}\,
  \frac{m_N}{36\,\pi}\,\Lambda\,k_F^{6} + I_{0F}^{\rm reg}(k_F) \, , \nonumber \\
  \label{eq:I0F-sum}
\end{eqnarray}
where the regular piece may be further subdivided into
\begin{eqnarray}
  && I_{0F}^{\rm reg}(k_F) =  I_{0Fa}(k_F) + I_{0Fb}(k_F) \, .
  \label{eq:I0F-reg-sum}
\end{eqnarray}
Their explicit evaluation yields
\begin{eqnarray}
  I_{0Fa}(k_F) &=&
  -\frac{1}{{(2\pi^2)}^2}\,\frac{8\,m_N}{315\,\pi^2}\,k_F^7 \, , \\
  I_{0Fb}(k_F) &=&
  -\frac{1}{{(2\pi^2)}^2}\,\frac{m_N}{630\,\pi^2}\,
  \left( -49 + 6\,\log{2}\right)\,k_F^7 \, ,  \nonumber \\
\end{eqnarray}
where by including them in Eqs.~(\ref{eq:I0F-reg-sum}) and (\ref{eq:I0F-sum}),
one recovers the result of Eq.~(\ref{eq:I0F-PDS}) for $I_{0F}(k_F; \Lambda)$.

\subsection{Higher order contacts}

As previously discussed, the loop correction to the energy per nucleon
can be written as
\begin{eqnarray}
  \frac{\delta E_2}{A} \propto
  2\,C_0 C_2\,I_{2F} + C_2^2\,I_{4F}^{(a)} + 2 C_0 C_4\,I_{4F}^{(b)} + \dots \, .
  \nonumber \\
\end{eqnarray}
For the density- and energy-dependent representations, the integrals
$I_{4F}^{(a)}$ and $I_{4F}^{(b)}$ are identical and the correction
simplifies to
\begin{eqnarray}
  \frac{\delta E_2}{A} \propto
  2\,C_0 C_2\,I_{2F} + \left( C_2^2 + 2 C_0 C_4 \right)\,I_{4F}  + \dots \, .
  \nonumber \\
\end{eqnarray}
Meanwhile, for the momentum-dependent representation it is convenient to
rewrite the correction as
\begin{eqnarray}
 && \frac{\delta E_2}{A} \propto 2\,C_0 C_2\,I_{2F}(k_F; \Lambda) +
  \left( C_2^2 + 2\,C_0 C_4 \right)\,I_{4F}(k_F; \Lambda) \nonumber \\ && \qquad
  + \left( C_2^2 - 2\,C_0 C_4 \right)\,\bar{I}_{4F}(k_F) \, ,
\end{eqnarray}
with $I_{4F} = (I_{4F}^{(a)} + I_{4F}^{(b)})/2$ and
$\bar{I}_{4F} = (I_{4F}^{(a)} - I_{4F}^{(b)})/2$, where in PDS
$\bar{I}_{4F}$ is independent of the regularization scale $\Lambda$.

For the density-dependent representation the $k_F$ dependence factorizes
and the loop integrals are trivial
\begin{eqnarray}
  I_{2F}(k_F; \Lambda) &=& k_F^2 I_{0F}(k_F; \Lambda) \, , \\
  I_{4F}(k_F; \Lambda) &=& k_F^4 I_{0F}(k_F; \Lambda) \, ,
\end{eqnarray}
where this pattern extends to higher orders.

For the energy dependent representation, before averaging in the Fermi sea
the loop integrals are still trivial and given by
\begin{eqnarray}
  I_{2F}(p, P, k_F; \Lambda) &=& p^2 I_{0F}(p, P, k_F; \Lambda) \, , \\
  I_{4F}(p, P, k_F; \Lambda) &=& p^4 I_{0F}(p, P, k_F; \Lambda) \, .
\end{eqnarray}
The averaging leads instead to
\begin{eqnarray}
  && I_{2F}(k_F; \Lambda) = -\frac{1}{{(2\pi^2)}^2}\,\Big[ \frac{m_N}{120\,\pi}\,\Lambda\,k_F^8 + \nonumber \\
    && \quad \frac{m_N}{5670\,\pi^2}\,(-79 + 24\,\log{2})\,k_F^9 \Big] \, ,
 \label{eq:I2F-energy} \\
  && I_{4F}(k_F; \Lambda) = -\frac{1}{{(2\pi^2)}^2}\,\Big[ \frac{m_N}{280\,\pi}\,\Lambda\,k_F^{10} + \nonumber \\
    && \quad \frac{m_N}{249480\,\pi^2}\,(-1337 + 564\,\log{2})\,k_F^{11} \Big] \, . \label{eq:I4F-energy}
\end{eqnarray}
Here it is interesting to notice that the divergent part is $\kappa_{2n}$ times
larger than the corresponding one in the density-dependent representation
(with the $\kappa_{2n}$ coefficients defined in Eq.~(\ref{eq:kappa-2n})).

For the momentum dependent representation it is convenient to define
the family of loop integrals
\begin{eqnarray}
  && I^{(2n,2m)}_F(p,P,k_F; \Lambda) = \nonumber \\ && \quad
  p^{2n}\,\int_{\Lambda} \frac{d^3 q}{(2 \pi)^3}\,q^{2m}\,
  \frac{\theta(q_1 - k_F)\,\theta(q_2-k_F)}{E - \frac{q^2}{m_N}} \, ,
  \nonumber \\ \\
  && I^{(2n,2m)}_F(k_F; \Lambda) = \nonumber \\ && \quad
  \int_{k_1, k_2 < k_F} \frac{d^3\,\vec{k}_1}{(2\pi)^3}\,
  \frac{d^3\,\vec{k}_2}{(2\pi)^3}\,\, I^{(2n,2m)}_F(p,P,k_F; \Lambda) \, .
  \nonumber \\
\end{eqnarray}
Note that $I_F^{(2,0)}(k_F; \Lambda)$ and $I_F^{(4,0)}(k_F; \Lambda)$ are simply
the energy-dependent loop integrals already calculated
in Eqs.~(\ref{eq:I2F-energy}) and (\ref{eq:I4F-energy}),
respectively.
The other three loop integrals of interest are
\begin{eqnarray}
  && I_{F}^{(0,2)}(k_F; \Lambda) = -\frac{1}{{(2\pi^2)}^2}\,\Big[ \frac{m_N}{120\,\pi}\,\Lambda\,k_F^8 + \nonumber \\
    && \qquad \frac{m_N}{5670\,\pi^2}\,(-255 + 24\,\log{2})\,k_F^9 \Big] \, ,
  \\
  && I_{F}^{(2,2)}(k_F; \Lambda) = -\frac{1}{{(2\pi^2)}^2}\,\Big[ \frac{m_N}{280\,\pi}\,\Lambda\,k_F^{10} + \nonumber \\
    && \qquad \frac{m_N}{249480\,\pi^2}\,(-3557 + 564\,\log{2})\,k_F^{11} \Big] \, , \\
  && I_{F}^{(0,4)}(k_F; \Lambda) = -\frac{1}{{(2\pi^2)}^2}\,\Big[ \frac{m_N}{280\,\pi}\,\Lambda\,k_F^{10} + \nonumber \\
    && \qquad \frac{m_N}{249480\,\pi^2}\,(-11321 + 564\,\log{2})\,k_F^{11} \Big] \, .
\end{eqnarray}
One can now calculate $I_{2F}$, $I_{4F}$ and $\bar{I}_{4F}$ in terms of the
previous integrals as
\begin{eqnarray}
  && \,\, I_{2F}(k_F; \Lambda) = 
  \frac{1}{2}\,I^{(2,0)}_F(k_F;\Lambda) +
  \nonumber \\
  && \qquad
  \frac{1}{2}\,I^{(0,2)}_F(k_F;\Lambda)
  \, ,  \\
  && \,\, I_{4F}(k_F; \Lambda) = \frac{3}{8}\,I^{(4,0)}_F(k_F;\Lambda) +
  \nonumber \\
  && \qquad
  \frac{1}{4}\,I^{(2,2)}_F(k_F;\Lambda) +
  \frac{3}{8}\,I^{(0,4)}_F(k_F;\Lambda) \, , \\
  && \,\, \bar{I}_{4F}(k_F; \Lambda) =  -\frac{1}{8}\,I^{(4,0)}_F(k_F;\Lambda) +
  \nonumber \\ && \qquad
  \frac{1}{4}\,I^{(2,2)}_F(k_F;\Lambda) 
  -\frac{1}{8}\,I^{(0,4)}_F(k_F;\Lambda) \, ,
\end{eqnarray}
where their explicit evaluation yields
\begin{eqnarray}
  && I_{2F}(k_F; \Lambda) =
  -\frac{1}{{(2\pi^2)}^2}\,\Big[ \frac{m_N}{120\,\pi}\,\Lambda\,k_F^8 +
    \nonumber \\
    && \qquad \frac{m_N}{5670\,\pi^2}\,(-167 + 24\,\log{2})\,k_F^9 \Big] \, ,
  \\
  && I_{4F}(k_F; \Lambda) = 
  -\frac{1}{{(2\pi^2)}^2}\,\Big[ \frac{m_N}{280\,\pi}\,\Lambda\,k_F^{10} +
    \nonumber \\
    && \qquad \frac{m_N}{249480\,\pi^2}\,(-5636 + 564\,\log{2})\,k_F^{11} \Big]
  \, , \label{eq:IF4-loop} \\
    && \bar{I}_{4F}(k_F) = 
  -\frac{1}{{(2\pi^2)}^2}\, 
  \frac{693\,m_N}{249480\,\pi^2}\,k_F^{11} \, .
  \label{eq:IF4-bar-loop}
\end{eqnarray}

Before finishing, it is interesting to notice that the momentum dependent
representation of the contact-range potential is ambiguous for operators
with four derivatives or more.
That is, there are multiple on-shell equivalent operators.
In the case of $C_4$ the two options are
\begin{eqnarray}
  C_4\,\frac{(p^4 + {p'}^4)}{2} \quad \mbox{or} \quad C_4 \,p^2 {p'}^2 \, ,
\end{eqnarray}
where for the second operator choice the loop integrals read
\begin{eqnarray}
  && I_{4F}'(k_F; \Lambda) = 
  -\frac{1}{{(2\pi^2)}^2}\,\Big[ \frac{m_N}{280\,\pi}\,\Lambda\,k_F^{10} +
    \nonumber \\
    && \qquad \frac{m_N}{249480\,\pi^2}\,(-4250 + 564\,\log{2})\,k_F^{11} \Big] \, , \\
    && \bar{I}_{4F}'(k_F) = 
  -\frac{1}{{(2\pi^2)}^2}\, 
    \frac{(-693)\,m_N}{249480\,\pi^2}\,k_F^{11} \, ,
\end{eqnarray}
which are labeled with a prima to distinguish them from the integrals
with the first choice, Eqs.~(\ref{eq:IF4-loop}) and (\ref{eq:IF4-bar-loop}).

\section{Power counting of the compensating terms}
\label{app:counting}

This Appendix explains how to derive the power counting of the lowest
order contact-range couplings within the formalism
of Section~\ref{sec:density-dependence}.
In it power counting is intimately intertwined with the power-law behavior of
the reduced wave function at short distances,
which determines the scaling of the contributions to the K- and G-matrices
from physics at distances below the (r-space) cutoff $R_c$:
\begin{eqnarray}
  && K(p) \Big|_{r < R_c} =
  4\pi\,\int_0^{R_c}\,dr\,V(r)\,\frac{\sin{p\,r}}{p}\,u_p(r) \, , \nonumber \\
  && G(p,P,k_F)
  \Big|_{r < R_c} =
  4\pi\,\int_0^{R_c}\,dr\,V(r)\,\frac{\sin{p\,r}}{p}\,u_{p,P}(r; k_F) \, .
  \nonumber \\
\end{eqnarray}
The arguments are identical in both cases and, for simplicity, only
the K-matrix will be discussed in detail.

If the two-body potential $V$ is regular, there will be a regular 
and irregular solution, where in the regular case
\begin{eqnarray}
  u_p^{\rm reg}(r) &=& \mathcal{N} \left[ r + \mathcal{O}(r^2) \right] \, ,
\end{eqnarray}
with $\mathcal{N}$ a normalization constant that does not
depend on the light scales of the system,
while in the irregular case
\begin{eqnarray}
  u_p^{\rm irr}(r) &=& \mathcal{N}(Q) \left[ 1 + \mathcal{O}(r) \right] \, , 
\end{eqnarray}
where the normalization constant is now a function of the light scales.
In this latter case
if the range of the potential $V$ is zero then a direct comparison
with the expected behavior of the reduced wave function for this type of
potential (that is, $u_p(r) = r - a_0$) yields
$\mathcal{N} = - a_0$, with $a_0$ the scattering length.
If the range of the potential is short (but not zero) then its effect
on the normalization constant is not necessarily trivial,
though it is sensible to assume  
\begin{eqnarray}
  \mathcal{N}(Q) \propto \frac{1}{Q} \, ,
\end{eqnarray}
just from its dimensionality alone.

From the previous arguments, in the regular case the zero momentum limit
of the $r < R_c$ contribution to the K-matrix is
\begin{eqnarray}
  v^{\rm reg}_{0}(R_c) = 4\pi\,\mathcal{N}\,\int_0^{R_c} dr\, V(r)\,\left[
    r^2 + \mathcal{O}(r^3) \right] \, ,
\end{eqnarray}
while in the irregular case it is:
\begin{eqnarray}
  v^{\rm irr}_{0}(R_c) = 4\pi\,\mathcal{N}(Q)\,\int_0^{R_c} dr\, V(r)\,\left[
    r + \mathcal{O}(r^2) \right] \, ,
  \nonumber \\
\end{eqnarray}
where $v_{0}$ is the first coefficient in the Taylor expansion of
the short-range K-matrix in terms of the momentum:
\begin{eqnarray}
  K(p) \Big|_{r < R_c} = \sum_{n=0}^{\infty} v_{2n}(R_c)\,p^{2n} \, . 
\end{eqnarray}
Provided the potential $V$ does not have a long-range component,
the only quantity containing a light scale in the previous
integrals is the normalization constant.
In particular, if the potential scales as
\begin{eqnarray}
  V(r) \propto M\,f(M r) \, ,
\end{eqnarray}
with $f(x) \to 0$ for $x \gg 1$ (to ensure a short-range potential) and
$f(x) \to f_0$ finite for $x \to 0$ (to simplify the analysis: otherwise
it will be necessary to take into account the explicit effect of
the potential in the scaling of the normalization
constant $\mathcal{N}$), one readily finds that $v_0$ scales as
\begin{eqnarray}
  v_0^{\rm reg}(R_c) &\propto& \mathcal{N} M R_c^3 \, , \\
  v_0^{\rm irr}(R_c) &\propto& \mathcal{N}(Q) M R_c^2 \, .
\end{eqnarray}
From the condition that the potential is short-ranged ($V(r) \sim 0$ for
$M r \gg 1$), this quantity saturates at $M R_c \sim 1$
\begin{eqnarray}
  \int_0^{R_c \gg 1/M} dr\, V(r)\, r^{\alpha}
  \approx \int_0^{R_c \sim 1/M} dr\, V(r)\, r^{\alpha} \, , \nonumber \\
\end{eqnarray}
where $\alpha$ is an exponent ($\alpha = 1$, $2$ for the irregular and
regular cases, respectively).
Thus one arrives at
\begin{eqnarray}
  v_0^{\rm reg}(R_c \gtrsim \frac{1}{M}) &\propto& \frac{1}{M^2} \, , \\
  v_0^{\rm irr}(R_c \gtrsim \frac{1}{M}) &\propto& \frac{1}{M Q} \, .
\end{eqnarray}
By matching these coefficients to a contact-range coupling, with
a delta-shell regulator at $r = R_c$ (with $M R_c \lesssim 1$),
one obtains the couplings
\begin{eqnarray}
  C_0^{\rm reg}(R_c) \propto \frac{1}{M^2} \quad \mbox{and} \quad
  C_0^{\rm irr}(R_c) \propto 
  \frac{R_c}{M} \, .
\end{eqnarray}
Then, by evolving into the infrared ($Q R_c \sim 1$),
one recovers the expected power counting
for $C_0^{\rm irr}$.

It is important to notice at this point that the previous power counting
argument is non-perturbative, and thus different in character to
the perturbative arguments commonly used in the literature.
Thus the previous argument can only reliably explain the infrared
enhancement of the ${\rm LO}$ coupling, which is non-perturbative.
If applied to the subleading couplings, it arrives to the conclusion that
\begin{eqnarray}
  C_{2n}(R_c) \sim \frac{1}{M^{2n}}\,C_0(R_c) \, .
\end{eqnarray}
Though correct in the regular case, for the irregular solution
this leads to the same $1/Q$ enhancement as with $C_0$,
instead of the usual $1/Q^2$ enhancement.
That is, the non-perturbative argument misses part of the infrared
enhancement of the subleading (i.e. perturbative) couplings.

This mismatch can be easily solved by considering the expansion of
the potential into a ${\rm LO}$ or non-perturbative part and
a subleading or perturbative correction
\begin{eqnarray}
  V &=& V_{\rm LO} + \delta V \, .
\end{eqnarray}
This entails a similar separation for the K-matrix and reduced wave function
\begin{eqnarray}
  K &=& K_{\rm LO} + \delta K \, , \\
  u_q &=& u_q^{LO} + \delta u_q \, ,
\end{eqnarray}
in which case the subleading correction to the K-matrix reads
\begin{eqnarray}
  \delta K(q) &=& 4\pi\, \int_0^{\infty} dr\,\delta V(r)\,u^{\rm LO}_q(r)\,u_q(r)
  \nonumber \\
  &=& 4\pi\, \int_0^{\infty} dr\,\delta V(r)\,{\left[ u^{\rm LO}_q(r) \right]}^2
  + \mathcal{O}\left( (\delta V)^2 \right) \, ,
  \nonumber \\
\end{eqnarray}
as can be deduced from distorted-wave perturbation theory
(see~\cite{PavonValderrama:2025azr} for a recent EFT-style exposition).
What changes now is that the (non-free) reduced wave function appears squared,
thus reproducing the RG arguments of Ref.~\cite{Valderrama:2014vra}.
If one integrates out again the short-range physics, the result
is that the coefficients $\delta v_{0}$ will be now enhanced
by a $\mathcal{N}^2$ factor,
\begin{eqnarray}
  \delta v_0^{\rm reg}(R_c) &\propto& \mathcal{N}^2 M R_c^3 \, , \\
  \delta v_0^{\rm irr}(R_c) &\propto& \mathcal{N}^2(Q) M R_c \, ,
\end{eqnarray}
leading to
\begin{eqnarray}
  \delta v_0^{\rm reg}(R_c \gtrsim \frac{1}{M}) &\propto& \frac{1}{M^2} \, , \\
  \delta v_0^{\rm irr}(R_c \gtrsim \frac{1}{M}) &\propto& \frac{1}{Q^2} \, ,
\end{eqnarray}
plus identical scalings for the $\delta v_{2n}(R_c)\,p^{2n}$ contributions.
When translated into the RG evolution of the subleading couplings,
this results in the usual $1/Q^2$ enhancement
in the irregular case.


%

\end{document}